\documentclass[%
 reprint,
 amsmath,amssymb,
 aps,
]{revtex4-1}

\usepackage{graphicx}
\usepackage{dcolumn}
\usepackage{bm}

\begin{document}

\preprint{APS/123-QED}

\title{Heralded single photon and correlated photon pair generation via spontaneous four-wave mixing in tapered optical fibers}

\author{A.A. Shukhin$^{1}$}
\email{E-mail: anatoly.shukhin@mail.ru}
\author{J. Keloth$^{2}$, K. Hakuta$^{2}$, and A.A. Kalachev$^{1}$}

\address{$^{1}$ Zavoisky Physical-Technical Institute, Kazan Scientific Center of the Russian Academy of Sciences, 10/7 Sibirsky Tract, Kazan 420029, Russia}
\address{$^{2}$ Center for Photonic Innovations, University of Electro-Communications, Chofu, Tokyo 182-8585, Japan}

\date{\today}

\begin{abstract}
We study the generation of correlated photon pairs and heralded single photons via strongly non-degenerate spontaneous four-wave mixing (SFWM) in a series of identical micro-/nano fibers (MNF). Joint spectral intensity of the biphoton field generated at the wavelength of about 880 nm and 1310 nm has been measured under excitation by 100 ps laser pulses demonstrating good agreement with the theoretical prediction. The measured zero-time second-order autocorrelation function was about 0.2 when the emission rate of the heralded photons was of 4 Hz. The MNF-based source perfectly matches standard single-mode fibers, which makes it compatible with the existing fiber communication networks. In addition, SFWM observation in a series of identical MNFs allows increasing generation rate of single photons via spatial multiplexing.
\end{abstract}

\maketitle

\section{Introduction}

Currently, much attention is being paid to developing non-classical light sources, such as those of single-photon and entangled two-photon states, which remains an important task in the field of quantum optical technologies \cite{o2009photonic,Flamini:2019bp}. One of the promising approaches to the problem is the use of nonlinear optical effects. Indeed, spontaneous parametric down-conversion and spontaneous four-wave mixing (SFWM) have been widely used for generation of entangled photon pairs and heralded single photons \cite{eisaman2011invited}. In particular, significant experimental progress has been achieved in implementing SFWM-based fiber sources that perfectly match the existing fiber communication networks \cite{wang2001generation, sharping2001observation, fiorentino2002all, takesue2004generation, li2005optical, rarity2005photonic, fan2005efficient, lin2006photon, lee2006generation}. In this respect, tapered optical fibers or micro-/nano-fibers (MNF) seem to be very promising nonlinear materials \cite{brambilla2010optical,tong2012optical,morrissey2013spectroscopy,balykin2014quantum,nayak2018nanofiber} that have a number of unique features due to the small mode diameter, significant evanescent field, small weight and size. In particular, MNF can be used for optical sensing \cite{chen2013review}, observation of nonlinear optical effects at very low pump power level (below 10 nW) \cite{spillane2008observation}, manipulating single atoms \cite{le2004atom}, and for efficient coupling between light and matter \cite{le2005spontaneous}. Subwavelength diameter of MNF allows one to construct miniature optical devices, which are characterized by small losses and low energy consumption, including miniature high-Q resonators \cite{brambilla2010optical}. Compared to the sources utilizing standard optical fibers, MNFs provide higher nonlinearity, which allows one to reduce the fiber length. In addition, the dispersion of a MNF can be tailored by controlling the fiber diameter, thereby separating spectra of SFWM and Raman scattering. At the same time, adiabatic fiber tapers are characterized by very small losses so that MNFs also perfectly match optical communication lines.

Generation of correlated two-photon and heralded single-photon states via SFWM in a MNF has been studied recently in \cite{cui2013generation,su2018micro,kim2019photon}. Correlated photon pairs were also generated in highly nonlinear chalcogenide tapered microwires \cite{meyer2015power}. In the present work, we demonstrate a broadband frequency-correlated photon pair source based on the strongly non-degenerate SFWM in a MNF. Compared to the previous experiments, we study joint spectral intensity of the biphoton field. In doing so, we observe SFWM in a series of identical MNFs, which can be used for increasing generation rate of heralded single photons via spatial multiplexing. 

\section{Micro-/nanofiber fabrication}

\begin{figure}[ht]
	\center{\includegraphics[clip,width=.5\textwidth]{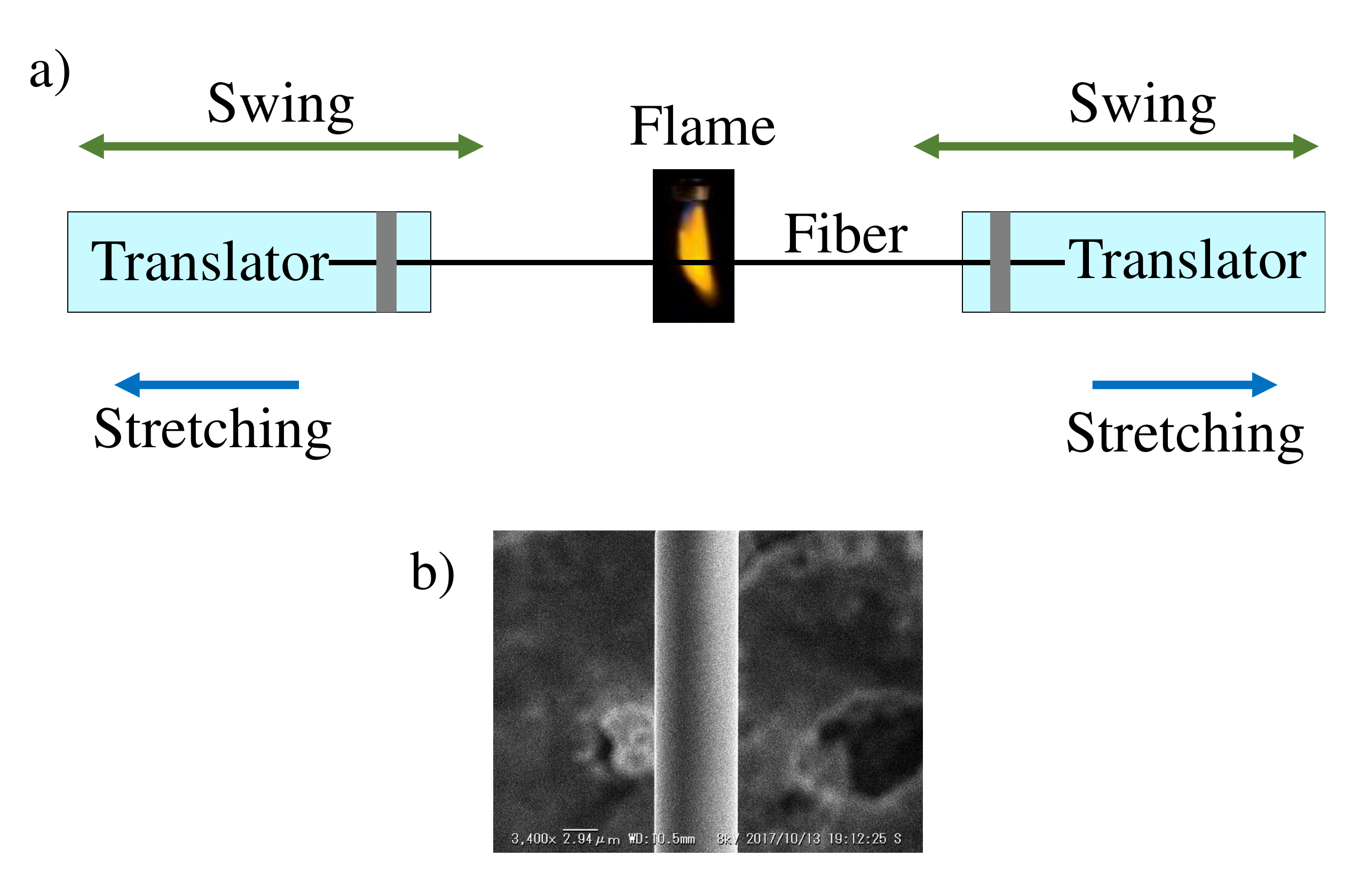}}
    \caption{Schematic representation of the MNF pulling setup (a), and electron microscope image of the fabricated MNF (b)}
    \label{MNFsetup}
\end{figure}

Subwavelength optical fibers with the cladding diameter less than $\sim 1\;\mu\rm{m}$ were made from the standard optical fibers ($d_{core} (\rm{Ge:SiO}_{2}) = 10\,\mu$m, $d_{cladding} (\rm{SiO}_{2}) = 125\,\mu$m) by using heating and stretching method \cite{tong2012optical}. MNF pulling setup is schematically shown in Fig.~\ref{MNFsetup}(a). 

\begin{figure}[ht]
     	\center{\includegraphics[clip,width=0.45\textwidth]{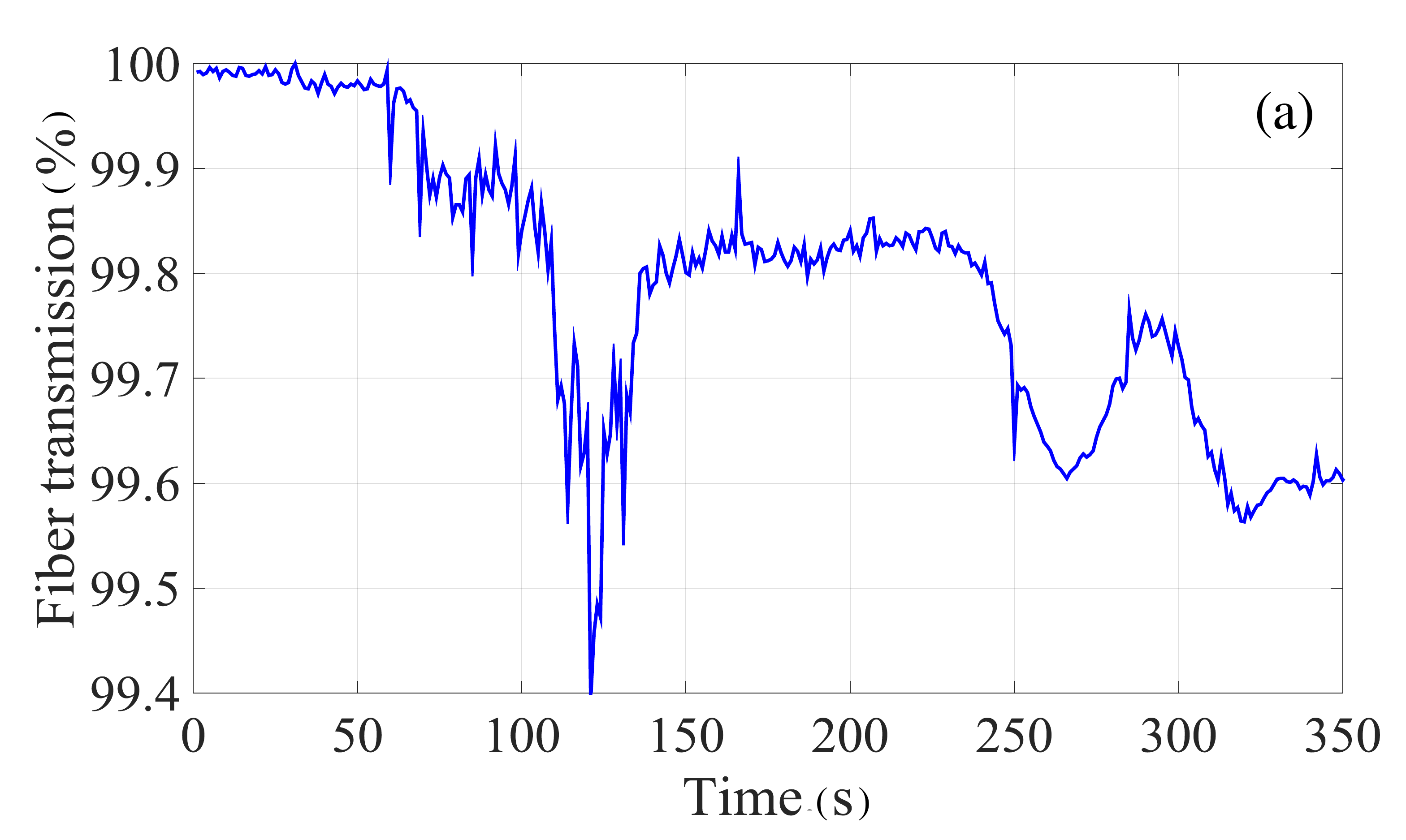}}
     	\center{\includegraphics[clip,width=0.45\textwidth]{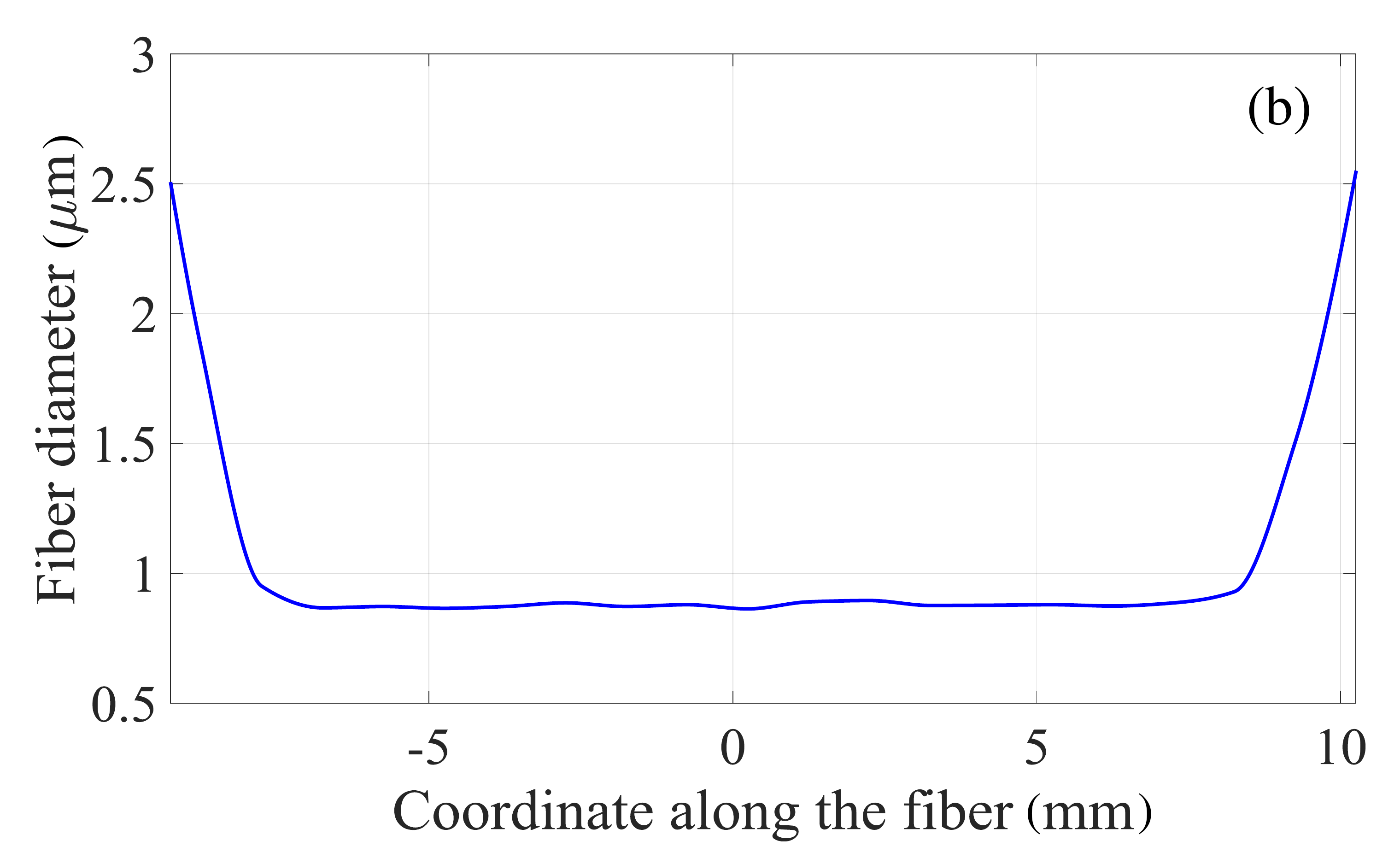}}
     	\label{profile}
    	\caption{Optical fiber transmission coefficient as a function of time while MNF pulling (a); the fiber diameter in the waist region as a function of the longitudinal coordinate (b)}
    	\label{transmission_and_profile}
\end{figure}

While tapering, the optical fiber is fixed on two moving translation stages and placed in front of motionless flame of burning O${}_2$ and H${}_2$ mixture. The movement of the translators consists of two types: synchronous displacement relative to the flame (swing) that determines heating zone and displacement in opposite directions (stretching) that determines fiber stretching distance and correspondingly MNF region. These movements are determined by four parameters: swing speed, swing distance, stretching speed and stretching distance so that a required adiabatic MNF profile can be achieved \cite{ward2014contributed} by adjusting them. As an example, an image of one of the fabricated MNFs is shown in Fig.~\ref{MNFsetup}(b). To control the MNF fabrication process, we measured the fiber transmission coefficient as a function of time while MNF pulling (Fig.~\ref{transmission_and_profile}(a)). This dependence has three specific regions: 1) multi-mode losses region ($50-70$~s); 2) core mode cut-off region ($70-290$~s) and 3) single-mode losses region ($290-350$~s). The resulting transmission of all the fabricated MNFs at the wavelength of $\lambda = 680$~nm is measured to be $T\approx 99.6$\%.

SFWM is a third-order optical nonlinear process, whereby the nonlinear interaction occasionally leads to the annihilation of two photons of the pump field and the simultaneous creation of two photons, which are usually called signal and idler, with different (nondegenerate regime) or the same (degenerate regime) frequencies. Due to isotropic symmetry of the nonlinear medium, created photons have the same polarization as those of the pump. 
The process of SFWM satisfies the following energy and phase-matching conditions:
\begin{equation}
2\omega_{p} = \omega_{s} + \omega_{i},
\end{equation}
\begin{equation}
2\vec{k}_p = \vec{k}_s + \vec{k}_i,
\end{equation}
where $\omega$ and $\vec{k}$ are the frequency and wave-vector, respectively, of the pump ($p$), signal ($s$) and idler ($i$) photons. We used pump radiation at the central wavelength of 1062~nm. With such pump, it is possible to observe strongly nondegenerate regime of SFWM so that the wavelength of the idler photon falls into the telecommunication band while that of the signal photon proves to be in the spectral range where detection efficiency is high enough. For this purpose, to fulfill conditions (1) and (2), the MNF waist diameter should be about 900~nm. In addition, for high efficiency of SFWM to be achieved, MNFs with a long waist region are required. In the present work, a series of eleven MNFs has been fabricated with the waist diameter of $D\approx 890\pm 12$~nm and the waist length of $L\approx 14$~mm. The result of MNF characterization by SEM is shown in Fig.~\ref{transmission_and_profile}(b). The fabricated MNFs have been sealed inside the plastic box to protect them from the dust.

\section{SFWM in an optical micro-/nanofiber}

\subsection{Theory}

A two-photon (biphoton) state of SFWM field generated in a MNF can be calculated in the first-order perturbation theory of quantum mechanics \cite{garay2010conversion}. Under this approach, the state vector of the biphoton field for a MNF of a variable cross-section can be written as follows: 
\begin{eqnarray}
|\psi\rangle = && |0\rangle  + A\int F(\omega_s,\omega_i) \times \nonumber \\
&& a_{m_s n_s}^\dagger (\omega_s)a_{m_i n_i}^\dagger (\omega_i) |0\rangle d\omega_s d\omega_s,
\end{eqnarray}
where $ |0\rangle $ corresponds to the vacuum state of the electromagnetic field, $A$ is a constant proportional to the nonlinearity of the fiber material, $a_{mn}^\dagger (\omega)$ is the photon creation operator in a fiber spatial mode with the distribution $u_{mn}(\rho)$, $\int |u(\rho)|^2 d^2\rho=1$, $d^2 \rho = dxdy$, and frequency $\omega$,
\begin{equation}
F(\omega_s,\omega_i)=\mathcal{I}(\omega_s,\omega_i)\,\mathcal{J}(\omega_s,\omega_i)
\end{equation}
is the joint spectral amplitude (JSA), which is proportional to the phase-matching function,
\begin{eqnarray}
\mathcal{J}(\omega_s,\omega_i)= && \sum_{q=1}^{N} \Bigg({\rm{sinc}}\Big(\frac{\Delta k_{q}(\omega_s,\omega_i)l}{2}\Big)\times\nonumber \\ 
&& \exp\Big(\frac{i\Delta k_{q}(\omega_s,\omega_i)l}{2}\Big) \times\nonumber \\
&& \exp\Big(i\sum_{n=q+1}^{N} \Delta k_{n}(\omega_s,\omega_i)l\Big) \times\nonumber \\
&& \eta_{m_p n_p,m_p n_p, m_s n_s, m_i n_i}^{q} \Bigg), 
\end{eqnarray}
and the pump function, $\mathcal{I}(\omega_s,\omega_i)$, that has the form of convolution of the pump field spectral amplitude,
\begin{equation}
\mathcal{I}(\omega_s,\omega_i)=\int d\omega_p E_p(\omega_p)E(\omega_i+\omega_s-\omega_p).
\end{equation}
Here, following \cite{Katsenelenbaum}, the MNF is divided by the $N$ cross-sections of the length $l$ each, so that the total phase-matching function is the sum of those calculated for the each cross-section. In doing so,
\begin{eqnarray}
\eta_{m_p n_p,m_p n_p, m_s n_s, m_i n_i}^{q} = &&\int u_{m_p n_p} (\rho) u_{m_p n_p} (\rho) \times\nonumber \\
&& u_{m_s n_s}^\ast (\rho) u_{m_i n_i}^\ast (\rho)\,
d^2 \rho
\end{eqnarray}
is the overlap integral between the four interacting modes that is calculated over $q$th MNF cross-section, 
\begin{equation}
\Delta k_q = k_q(\omega_p)+k_q(\omega_i+\omega_s-\omega_p)-k_q(\omega_s)-k_q(\omega_i)
\end{equation}
is the wave vector mismatch for $q$th MNF cross-section, $k_q(\omega)=\omega n_{eff,q}(\omega)/c$, $n_{eff,q}(\omega)$ is the effective refractive index,
\begin{equation}
E_p(\omega_p)=E_{p_0}\exp\Big(-\frac{(\omega_p-\omega_{p_0})^2}{2\sigma^2}\Big)
\end{equation}
is the spectral amplitude of the pump field, which corresponds to a Gaussian pulse with the duration of $\tau_p=2\sqrt{\ln 2}/\sigma$. Eq.~(5) is written for the case of a MNF with a variable waist diameter, which generalizes results obtained in \cite{garay2010conversion} for a MNF of a constant diameter. It should be noted that each cross-section is characterized by its own diameter, overlap integral, wave vector mismatch, and additional phase (the second exponent in Eq.~(5)) acquired by light due to the propagation through the rest of the cross-sections.

\begin{figure}[ht]
     	\center\includegraphics[clip,width=0.5\textwidth]{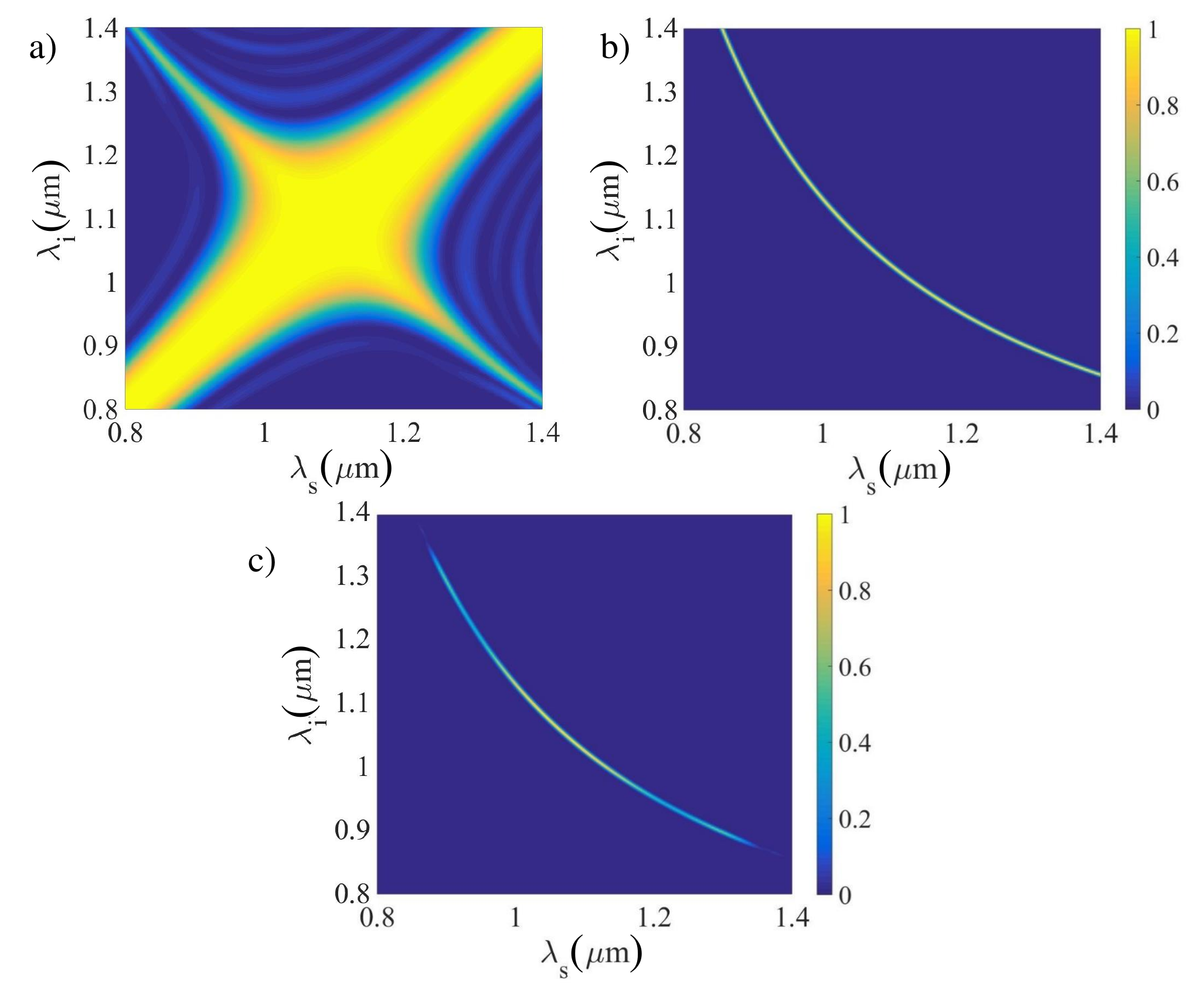}
    	\caption{Theoretically calculated phase-matching function (a), pump function (b) and joint spectral intensity (c) for the MNF which profile is shown in Fig.~\ref{transmission_and_profile}(b)}
	    \label{JSIt}
\end{figure}

Due to the small diameter of the MNF, the average value of the overlap integral (7) in our case is equal to $1\cdot 10^{12}\;\rm{m}^{-2}$ for a MNF waist diameter of $890$~nm, which is larger than that for a standard optical fiber by two orders of magnitude. The phase-matching function (5) is calculated taking into account the measured profile of the fabricated MNFs (Fig.~\ref{transmission_and_profile}(b)). The squared absolute value of this function shown in Fig.~\ref{JSIt}(a) describes probability density of the photon pair generation as a function of the wavelengths in the limit of an infinitely broad pump field. Fig.~\ref{JSIt}(b) shows the spectral function (7) corresponding to the pump field. As such, we used the radiation of a pulsed laser consisting of the Fianium master oscillator and the ytterbium-doped fiber-based amplifier (central wavelength 1062~nm, pulse width 100~ps, pulse repetition rate 18~MHz). Since the pump pulses are not bandwidth-limited, the spectral width of the pump field is about 2~nm in our case (instead of $10^{-3}$~nm for bandwidth-limited-$100$~ps pulses).

Multiplication of the spectral pump function and phase-matching function corresponds to the JSA of the biphoton field (5). Square of the module of this function (joint spectral intensity) is shown in Fig.~\ref{JSIt}(c). From this figure we can see that there is an opportunity of photon pair generation in the non-degenerate regime, where the idler photons are generated in the telecommunication O-band, while the signal ones are generated in the spectral range that is close to the visible light where detection efficiency is high enough.

Joint spectral intensity (JSI) reveals the most important spectral properties of the biphoton field. Particularly, besides the spectral width and central wavelength of the generated photons, it allows one to analyze frequency correlations, which is important for heralded preparation of single-photon states \cite{mosley2008heralded,spring2013chip}. Thus, measuring JSI for a fabricated MNF is important step for characterization of the nonclassical light source. 

\subsection{Experiment}

\begin{figure}[ht]
     	\center\includegraphics[clip,width=0.5\textwidth]{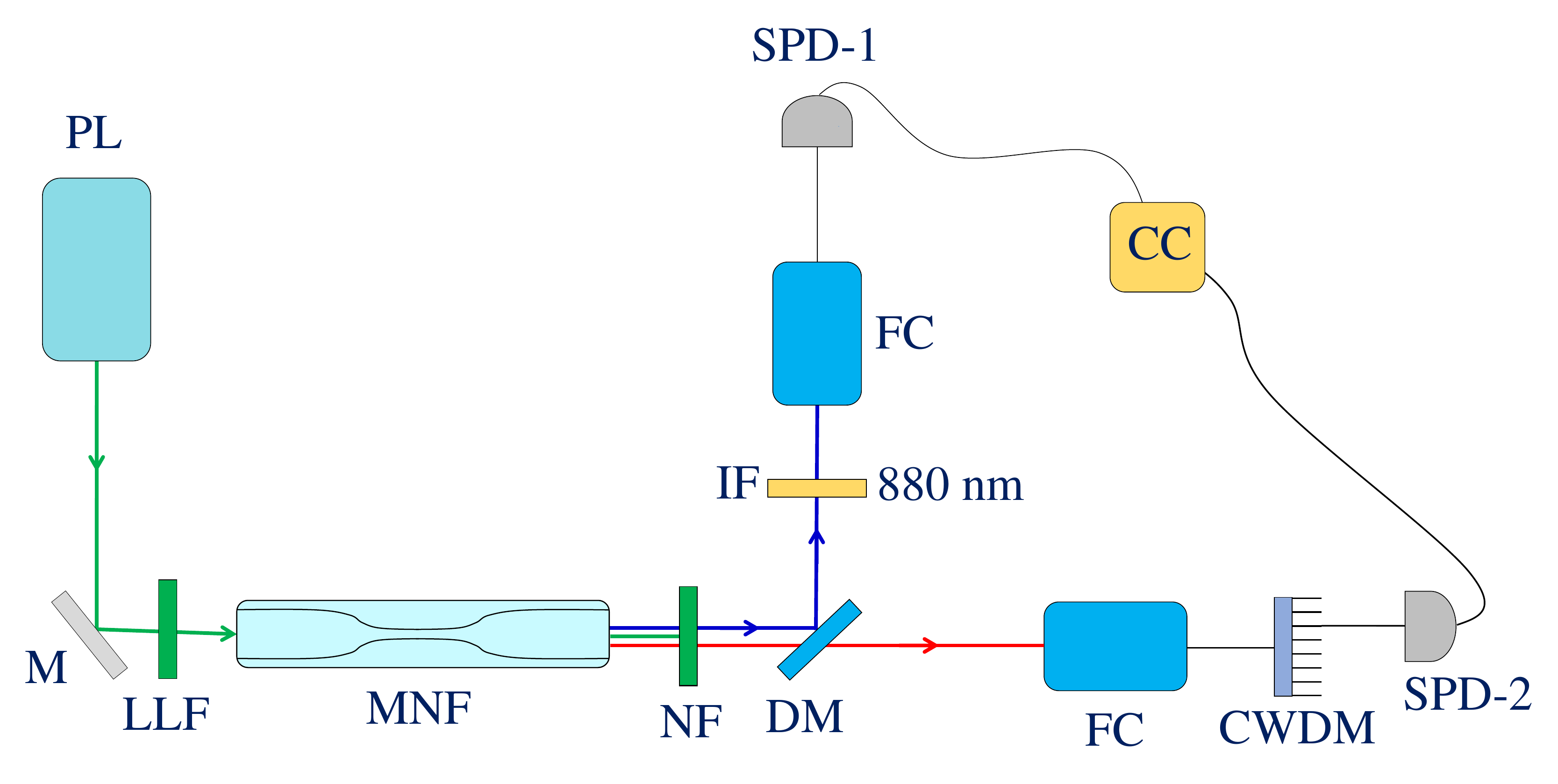}
    	\caption{Schematic of the experimental setup for studying SFWM in optical MNFs. PL is the pump laser, M is the mirror, LLF is the laser line filter, NF is the notch filter, DM is the dichroic mirror, IF is the interference filter, FC is the free space to fiber coupler, SPD is the single-photon detector, CC is the coincidence counter}
	    \label{expsetup}
\end{figure}

The experimental setup is schematically shown in Fig.~\ref{expsetup}. 
The pump field passed through a laser line filter (LLF) and was focused with a lens into the tapered single-mode fibers. To cut off the pump field radiation after the fiber we used four notch filters (NF) with 40 dB attenuation of each at the pump wavelength. The biphoton field generated in the MNF at the wavelengths about 880~nm (signal) and 1310~nm (idler) was divided by using a dichroic mirror (DM). Idler photons have been filtered by using a coarse wavelength division multiplexing unit (CWDM) which has 18 by pass channels covering spectral range from 1270 nm to 1610 nm with 20-nm bandwidth (FWHM) of each. After passing through one of these channels, idler photons went to the IR range single-photon detector SPD-2 (ID210, ID Quantique). Signal photons have been filtered by using an interference filter (IF) with the central wavelength of 880~nm and the bandwidth of 40~nm. After passing through the filter signal photons went to the visible range single-photon detector SPD-1 (SPCM-AQRH, Perkin-Elmer). The detection quantum efficiencies of SPD-1 and SPD-2 are about 40\% and 12\% at the signal and idler wavelengths, respectively. Signals from the both SPDs have been compared on a coincidence counter CC (id800, ID Quantique) with the time resolution of 81~ps. It should be noted that the total attenuation of the pump was equal to $-220$~dB in the signal channel and $-200$~dB in the idler channel.

\begin{figure}[ht]
     	\center\includegraphics[clip,width=0.35\textwidth]{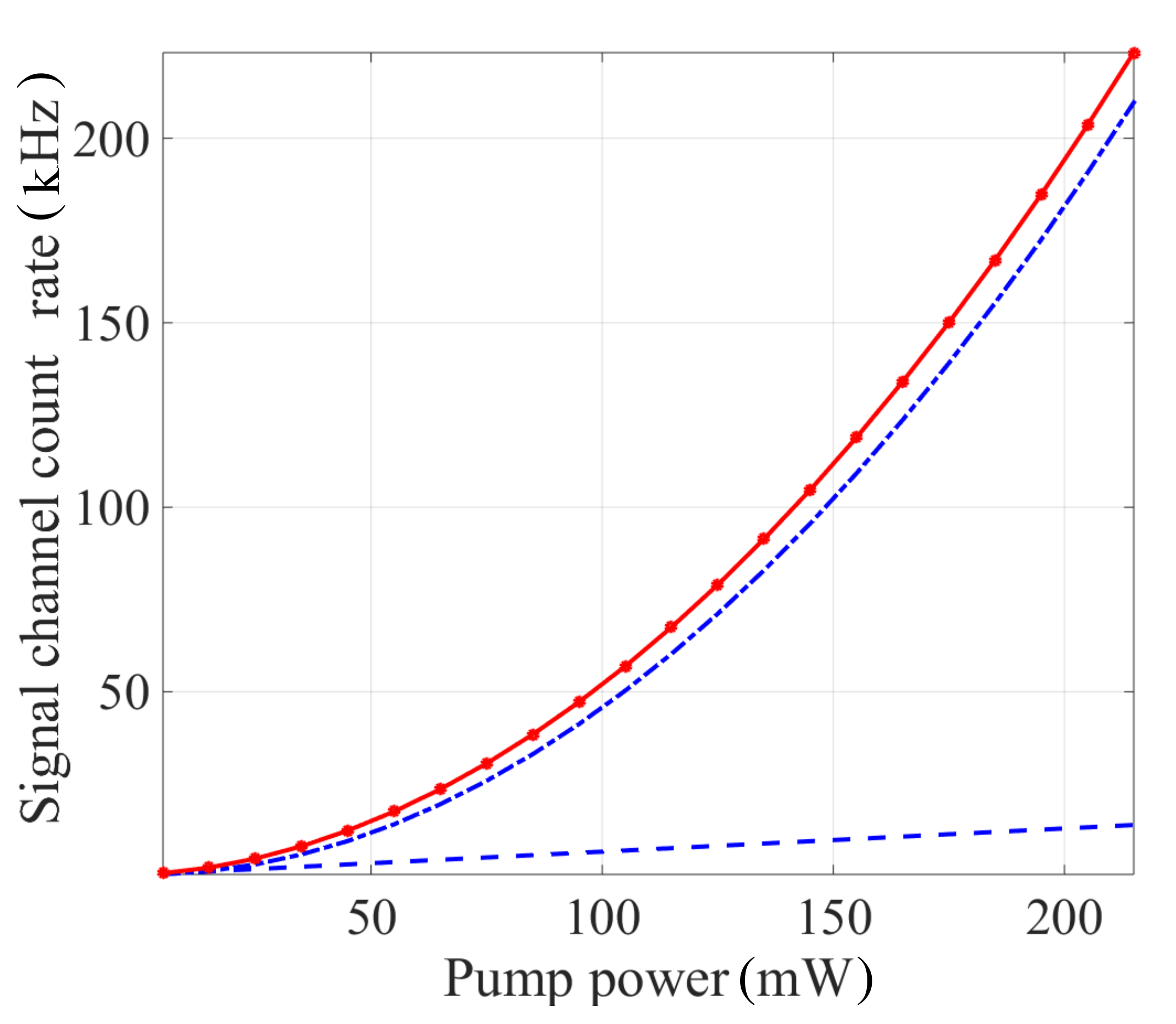}
    	\caption{Single count rate in the signal channel as a function of the average pump power. Red curve is the fitting of the experimental points by the second-order polynomial function $D+aP^2+bP$, blue dashed line and blue dot dashed curve is the linear ($b=0.057$~kHz/mW) and quadratic ($a=0.0043$~kHz/mW$^2$) terms of the fitting function, respectively. $D$ is the dark count rate (400 Hz).}
	    \label{signalrate}
\end{figure}

The photo-counts in both channels may be caused not only by SFWM, but also by Raman scattering of the pump field. To distinguish these contributions, we measured the single count rate in the signal channel as a function of the average pump power (Fig.~\ref{signalrate}) and fitted it by the second-order polynomial function $D+aP^2+bP$. 
Quadratic term in this function corresponds to the contribution of the photons created via SFWM, while the linear one represents Raman scattering \cite{cui2013generation, su2018micro}. The former proves to be close enough to the experimental points, which is the evidence of the high contribution of the SFWM signal.

\begin{figure}[ht]
     	\center{\includegraphics[clip,width=0.4\textwidth]{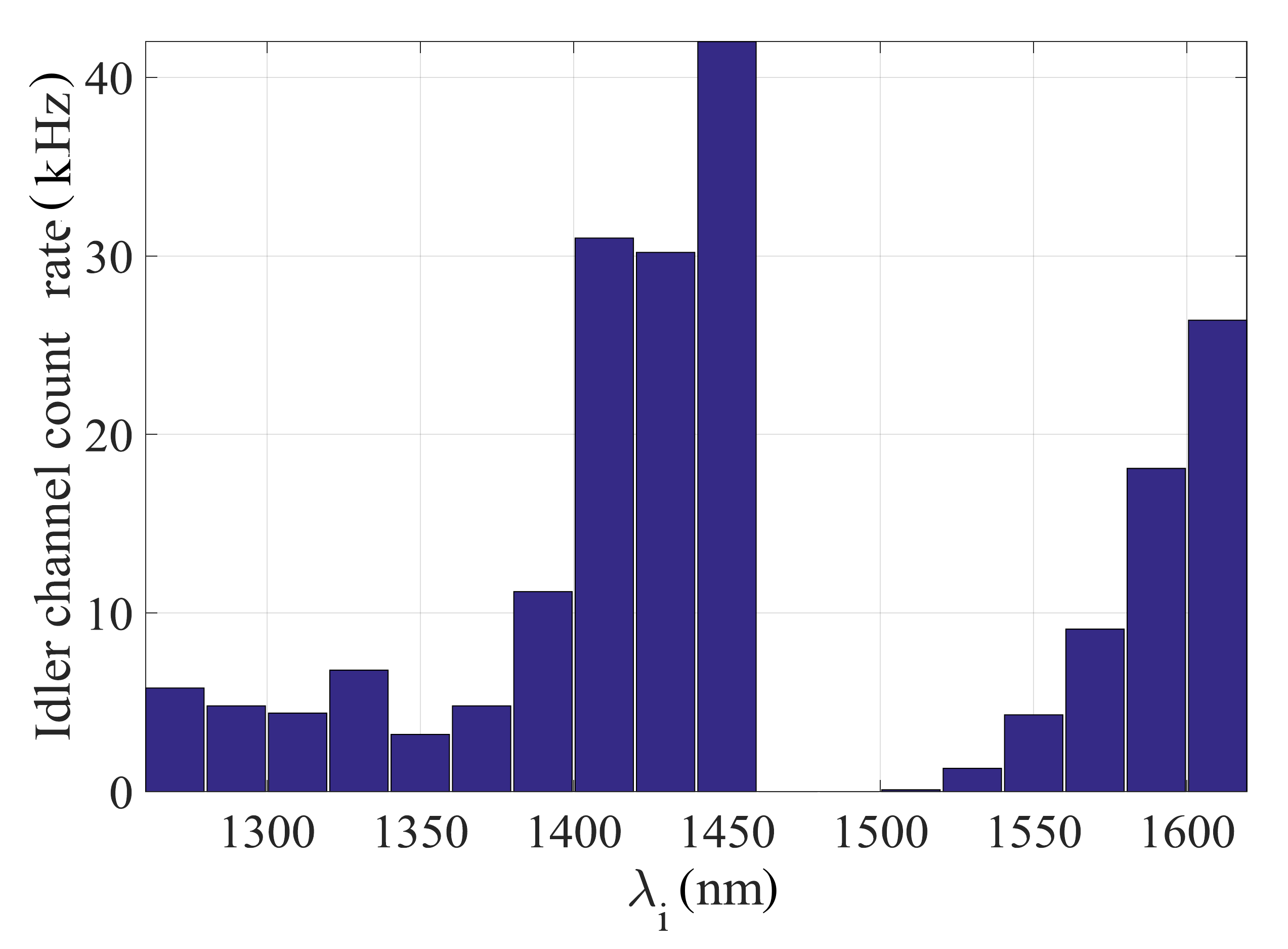}}
    	\caption{Single count rate in the idler channel as a function of wavelength (after subtracting background noise, which is equal to 15.2 kHz). Average pump power is of 118 mW}
    	\label{irate}
\end{figure}

We also measured the single count rate in the idler channel as a function of the wavelength by using CWDM (Fig.~\ref{irate}). Two peaks of intensity are observed at the wavelength of $1450$~nm and $1610$~nm. However, as will be shown below, the coincidence count rate demonstrates no signals for these CWDM channels, but only for the range of $1270-1350$~nm. Moreover, the photo-count rate in the idler channel depends linearly on the pump power, so that we can conclude that most of the photons in the idler channel appear due to the linear optical effects such as spontaneous Raman scattering. In particular, the intensity peak in the region of $1400-1450$~nm is the 5th-order Stokes line of Raman scattering in silica \cite{agrawal2012nonlinear}, while the peak around 1600~nm is interprited as an unidentified line \cite{cui2013generation}.

\begin{figure}[ht]
     	\center\includegraphics[clip,width=0.45\textwidth]{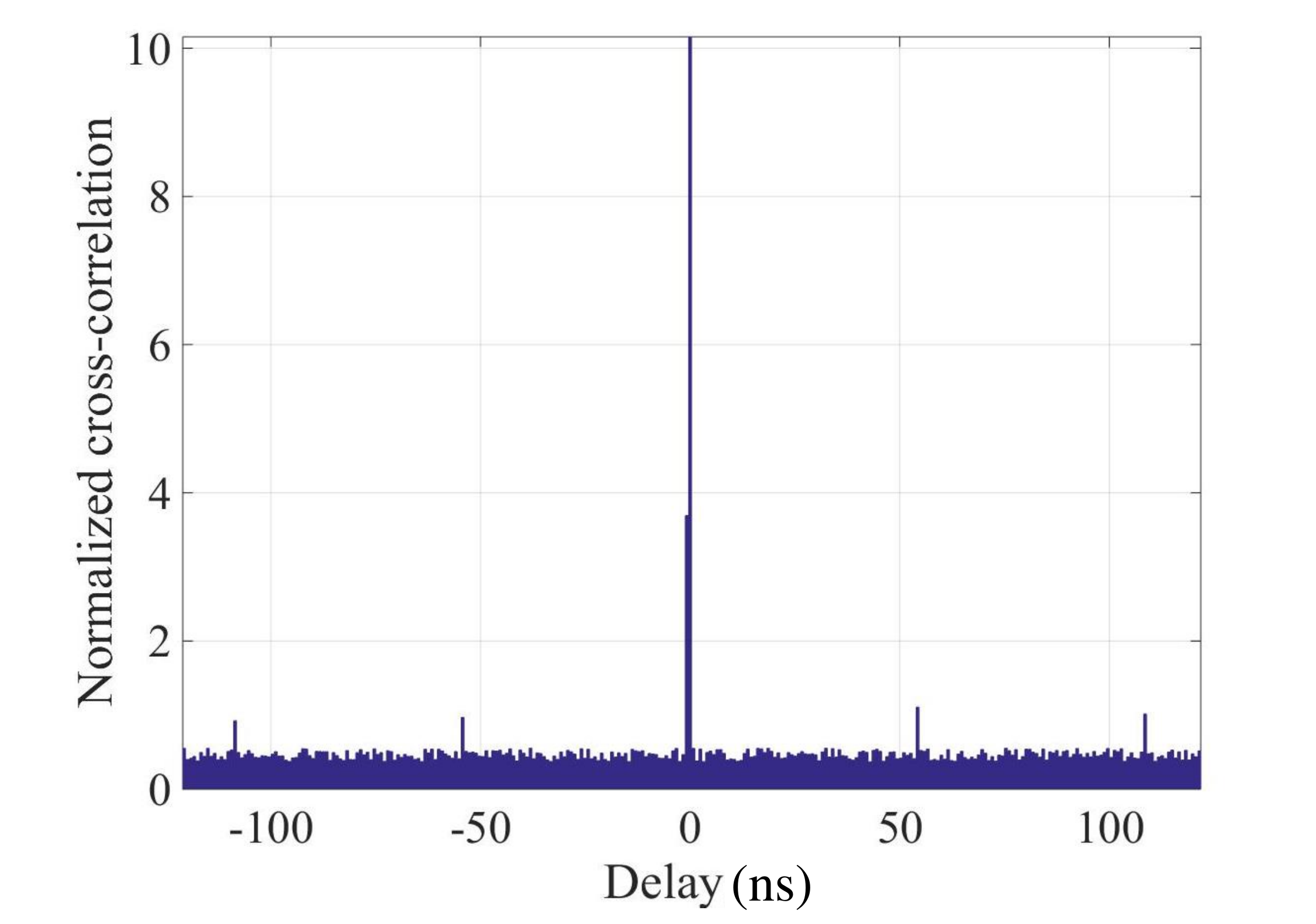}
    	\caption{Coincidence count rate as a function of time delay between the signal and idler photons. The width of each bar is of $810$~ps. The distance between the peaks corresponds to the time interval between the pump pulses and equals to $54$~ns. Average pump power is of $118$~mW}
	    \label{CC}
\end{figure}

\begin{figure}[ht]
     	\center\includegraphics[clip,scale=0.3]{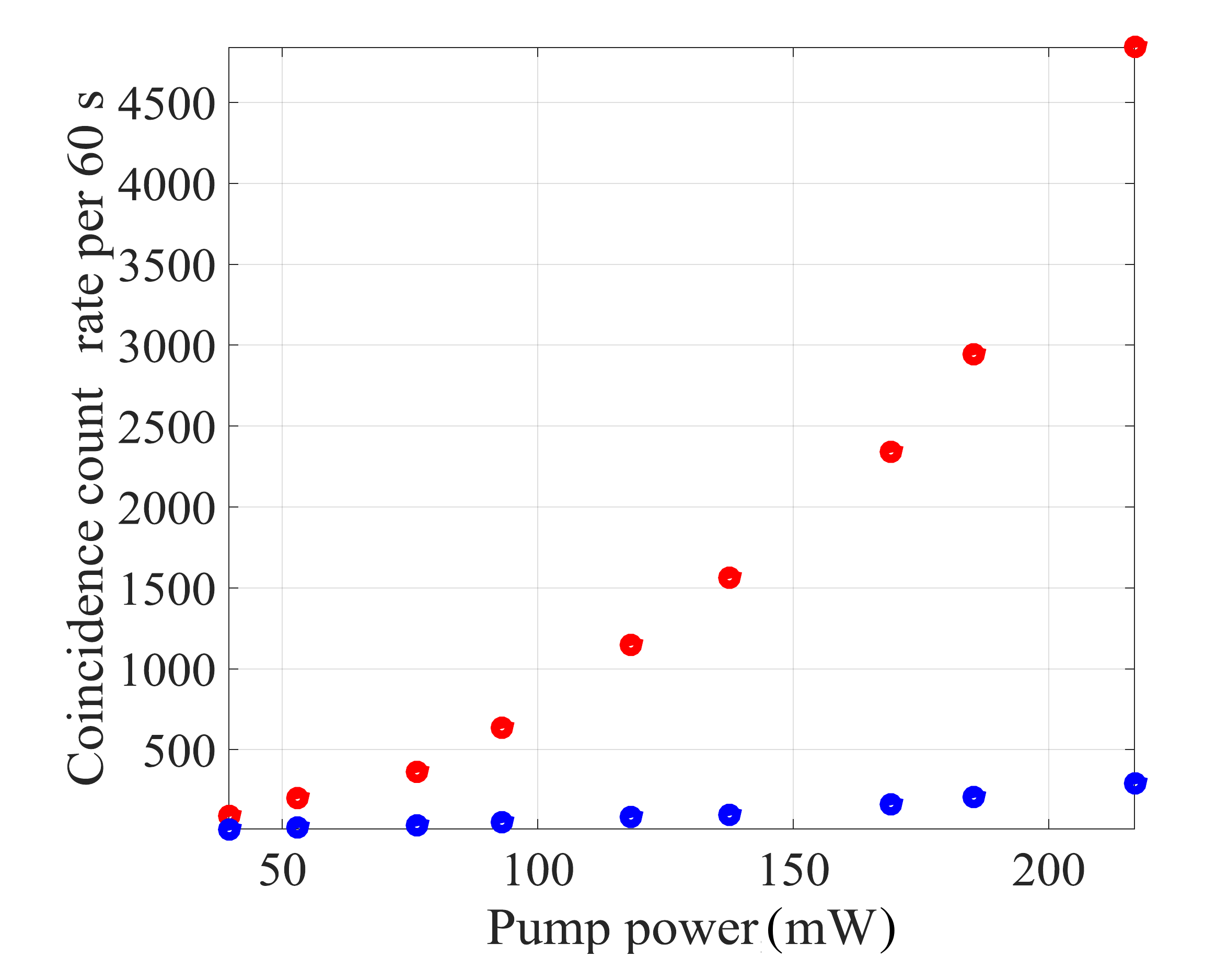}
    	\caption{Coincidence count rate (red dots) and accidental coincidence count rate (blue dots) as functions of the average pump power. CWDM channel central wavelength is of 1310 nm.}
	    \label{coinvspp}
\end{figure}

Fig.~\ref{CC} represents the normalized coincidence count rate as a function of the time delay between the signal and idler photons. The peak at the zero delay time corresponds to the strong temporal correlation between the photon detection events, which proves the two-photon nature of the generated light. The peak coincidence count rate and the accidental coincidence count rate as functions of the average pump power are shown in Fig.~\ref{coinvspp}. The observed quadratic dependence for the peak coincidence count rate reveals high signal-to-noise ratio.

\begin{figure}[ht]
     	\center\includegraphics[clip,width=0.5\textwidth]{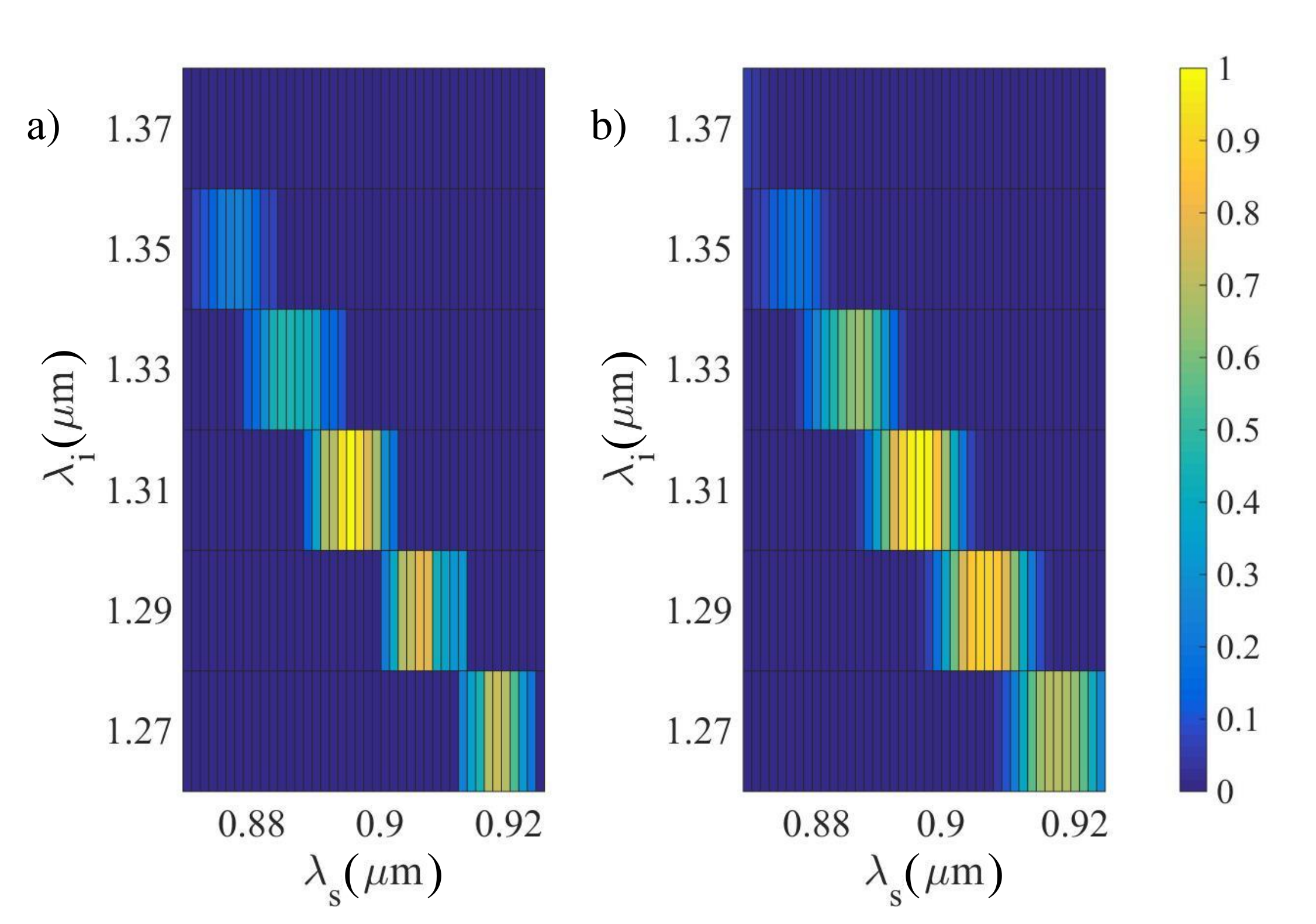}
    	\caption{Measured (a) and calculated (b) JSI for a given MNF. The colorbar represents the relative probability of observing correlated photon pairs. The average diameter of the MNF is estimated to be $900$~nm}
	    \label{JSI}
\end{figure}

The use of the variable spectral filtering in the signal and idler channels allowed us to measure the JSI of the biphoton field. For spectral photon selection we used a diffraction grating (100 lines/mm) in the signal channel (instead of $880$-nm interference filter) and the CWDM unit in the idler channel. By changing CWDM channels and rotating the diffraction grating we plotted the whole JSI function distribution in the experimentally accessible spectral range (Fig.~\ref{JSI}(a)). The resolution of the measurement was of $20$~nm in the idler field and $1.5$~nm in the signal field. Numerically calculated JSI demonstrates good agreement between the theory and experiment (Fig.~\ref{JSI}(b)). We also measured the JSI in different micro-/nanofibers and got approximately the same position of its peak. Comparing the measured JSIs with numerically calculated one allows us to estimate the diameter of the MNFs with accuracy of 3 nm. As a result, the waist diameter of all the MNFs lays within the $890\pm12$~nm region, as should be expected.

Following the papers \cite{garay2011spontaneous,garay2010conversion}, we calculated the photon pair generation efficiency $\eta$ for our MNFs and obtained $\eta=7\cdot 10^{-10}$ which gives the internal photon pair generation rate of $R_{internal}= 3\cdot 10^6$ pairs per second. The latter was found from the conversion efficiency by taking into account the number of photons in the pump pulse ($2.67\cdot 10^8$ at average pump power of $118$~mW) and the pump pulses repetition rate. Then by accounting the losses in the setup (which include efficiencies of SPDs and losses on the optical elements per each channel, in total $-17$~dB), the observed photon flux turns out to be $R_{observed} = R_{internal}\eta_i\eta_s \approx 60$ pairs per second (without spectral selection in the channels), which agrees well with the experimentally observed value of $40$ photon pairs per second for the average pump power of $118$~mW. Spectral filtering by using diffraction grating in the signal channel and CWDM unit in the idler channel reduces the observed photon flux down to 4 pairs per second.

\begin{figure}[ht]
     	\center\includegraphics[clip,width=0.5\textwidth]{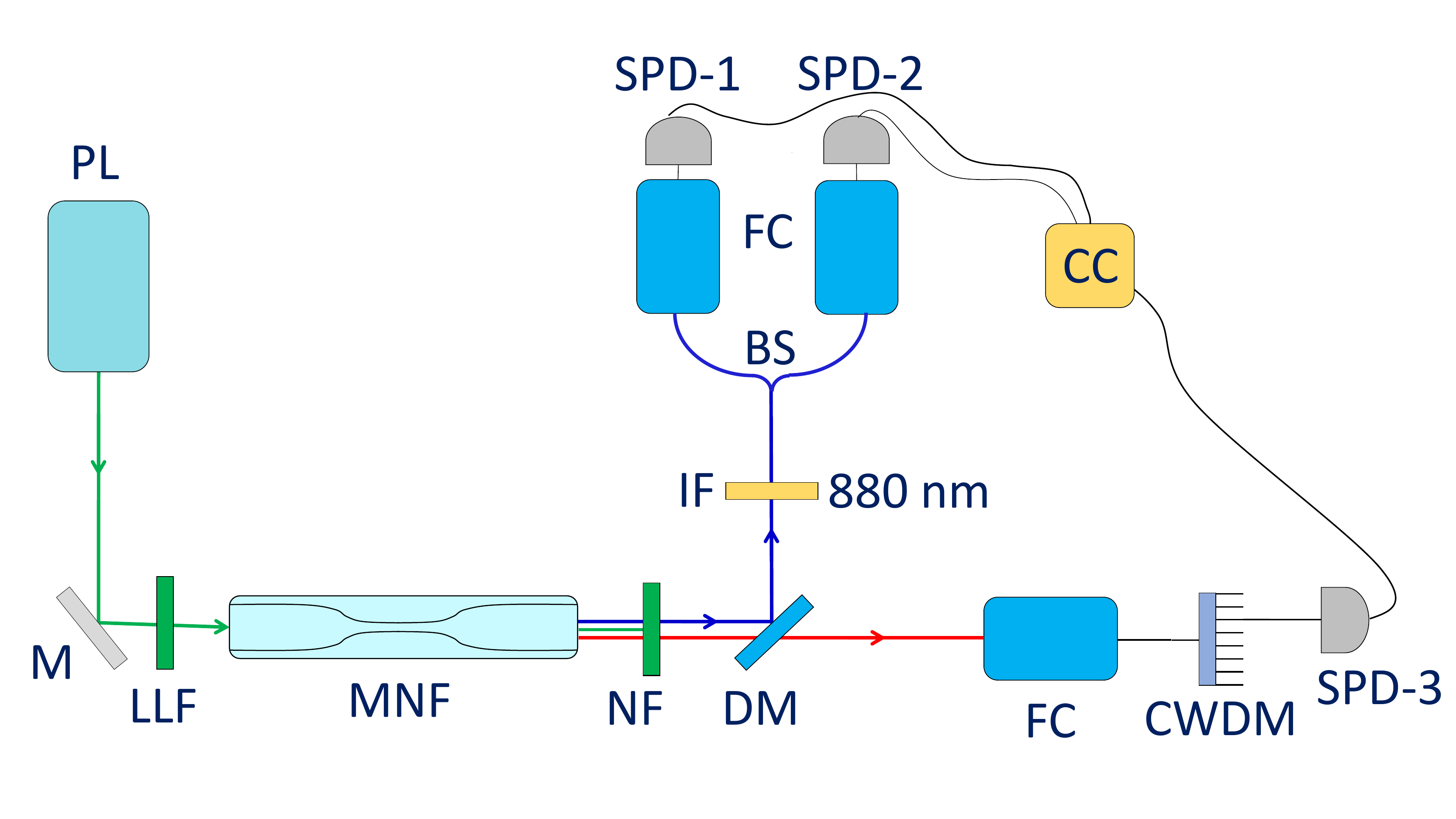}
    	\caption{Experimental setup for measuring heralded second-order auto-correlation. Here BS is a $50/50$ (47/53) beam splitter at $850~\rm{nm}$ ($880~\rm{nm}$)}
	    \label{setup2}
\end{figure}

\begin{figure}[ht]
     	\center\includegraphics[clip,width=0.5\textwidth]{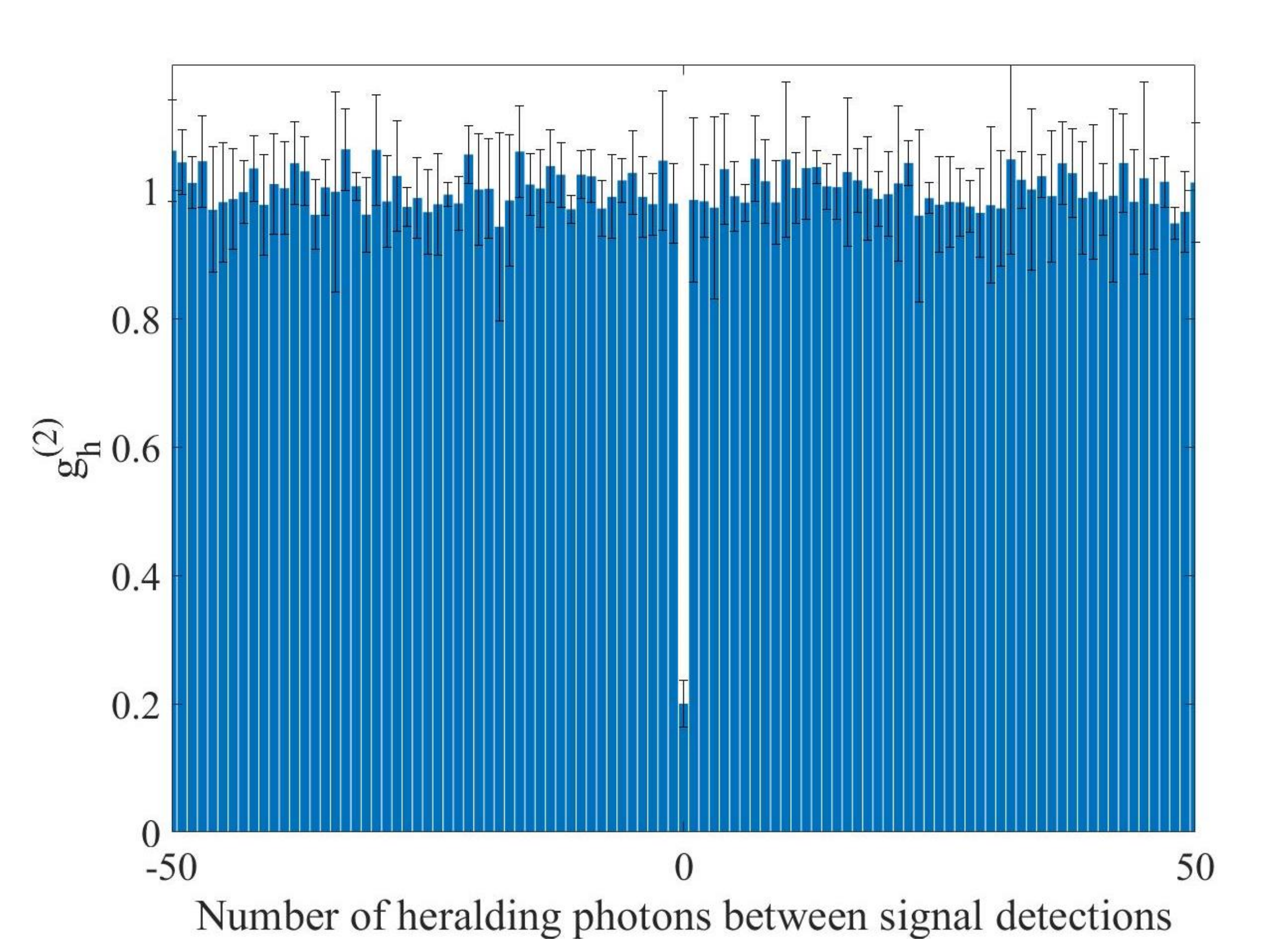}
    	\caption{Heralded auto-correlation histogram}
	    \label{hist}
\end{figure}

The fact of simultaneous pair creation via SFWM, allows one to use the source of correlated photon pairs for heralded single-photon generation, in which one photon from the pair is detected to herald the other. To study our source as a heralded single-photon one, we place a $50/50$ beam splitter in the signal (heralded) channel (Fig.~\ref{setup2}) and measured heralded auto-correlation histogram (Fig.~\ref{hist}).
The photon pairs (and therefore single photons) are created at random times and with a low probability due to the spontaneous third-order non-linear process. The heralded photo-detection events are therefore absolutely random and rare in comparison with the coincidence window. The latter was set to be of $810$~ps for covering the jitter of the detection system. For this reason, the measuring of the second-order auto-correlation function, where the argument is the time delay between two heralded photons, is not possible. Instead, following technique suggested in \cite{fasel2004high} and widely used for single-photon sources characterisation \cite{rielander2016cavity,seri2019quantum,seri2018laser,lenhard2017coherence,bock2016highly}, we measured heralded auto-correlation histogram, where the argument is the number of heralding photons between the nearest heralded ones detected either by SPD-1 or SPD-2. This technique allows one to reduce the long time intervals between heralded photons but preserves all the probabilities of single-photon and many-photon events. 
Experimentally, from the coincidence counting system we get the sequences ${t1}$, ${t2}$, and ${t3}$ of the time tags of the photons detected by SPD-1, SPD-2 and SPD-3, respectively. By analysing these time tags, we obtained the histogram shown in Fig.~\ref{hist}.
The anti-bunching drop at zero point indicates clearly the quantum behavior of the signal field and its single-photon statistics. For the average pump power of $118$~mW and generation rate of $4$~Hz, the value of zero-time second-order auto-correlation function (which is the same as the value of the zero-point auto-correlation histogram) is observed to be $g^{(2)}_{h}(0) \approx 0.2$.
Each of the bars is obtained by accounting all the photons detected within the coincidence window.
Low generation efficiency and large deadtimes ($\approx15~\mu$s) leaded to about the week of running experiment to accumulate the large enough number of three-fold coincidences and present statistically significant results.

\section{Conclusion}
In this work, we have fabricated a bunch of identical 14 mm long low-loss micro-/nanofibers and have studied spontaneous four-wave mixing in them by measuring coincidence counting rate as a function of the pump power and the wavelength of the generated photons. The measured joint spectral intensity of the biphoton field agrees well with the theoretically expected distribution. We also measured anti-bunching for the signal field, which is the evidence of its single-photon nature. The high transmittance of the fabricated MNFs potentially allows one to significantly increase the photon pair conversion efficiency by making an all-fiber resonator using, e.g., two Bragg gratings enclosing the nanofiber waist \cite{nayak2018nanofiber}.

\begin{acknowledgments}
A.A.S. thanks the Center for Photonic Innovations, University of Electro-Communications for hospitality. J.K. and K.H. thank the Japan Science and Technology Agency for the funding support through Strategic Innovation Program (Grant No. JPMJSV0918). The results of Sec.~III were obtained within the government assignment for FRC Kazan Scientific center of RAS.
\end{acknowledgments}

\nocite{*}

\bibliography{apssamp}

\providecommand{\noopsort}[1]{}\providecommand{\singleletter}[1]{#1}%
\begin{thebibliography}{38}%
\makeatletter
\providecommand \@ifxundefined [1]{%
 \@ifx{#1\undefined}
}%
\providecommand \@ifnum [1]{%
 \ifnum #1\expandafter \@firstoftwo
 \else \expandafter \@secondoftwo
 \fi
}%
\providecommand \@ifx [1]{%
 \ifx #1\expandafter \@firstoftwo
 \else \expandafter \@secondoftwo
 \fi
}%
\providecommand \natexlab [1]{#1}%
\providecommand \enquote  [1]{``#1''}%
\providecommand \bibnamefont  [1]{#1}%
\providecommand \bibfnamefont [1]{#1}%
\providecommand \citenamefont [1]{#1}%
\providecommand \href@noop [0]{\@secondoftwo}%
\providecommand \href [0]{\begingroup \@sanitize@url \@href}%
\providecommand \@href[1]{\@@startlink{#1}\@@href}%
\providecommand \@@href[1]{\endgroup#1\@@endlink}%
\providecommand \@sanitize@url [0]{\catcode `\\12\catcode `\$12\catcode
  `\&12\catcode `\#12\catcode `\^12\catcode `\_12\catcode `\%12\relax}%
\providecommand \@@startlink[1]{}%
\providecommand \@@endlink[0]{}%
\providecommand \url  [0]{\begingroup\@sanitize@url \@url }%
\providecommand \@url [1]{\endgroup\@href {#1}{\urlprefix }}%
\providecommand \urlprefix  [0]{URL }%
\providecommand \Eprint [0]{\href }%
\providecommand \doibase [0]{https://doi.org/}%
\providecommand \selectlanguage [0]{\@gobble}%
\providecommand \bibinfo  [0]{\@secondoftwo}%
\providecommand \bibfield  [0]{\@secondoftwo}%
\providecommand \translation [1]{[#1]}%
\providecommand \BibitemOpen [0]{}%
\providecommand \bibitemStop [0]{}%
\providecommand \bibitemNoStop [0]{.\EOS\space}%
\providecommand \EOS [0]{\spacefactor3000\relax}%
\providecommand \BibitemShut  [1]{\csname bibitem#1\endcsname}%
\let\auto@bib@innerbib\@empty
\bibitem [{\citenamefont {O'brien}\ \emph {et~al.}(2009)\citenamefont
  {O'brien}, \citenamefont {Furusawa},\ and\ \citenamefont
  {Vu{\v{c}}kovi{\'c}}}]{o2009photonic}%
  \BibitemOpen
  \bibfield  {author} {\bibinfo {author} {\bibfnamefont {J.~L.}\ \bibnamefont
  {O'brien}}, \bibinfo {author} {\bibfnamefont {A.}~\bibnamefont {Furusawa}},\
  and\ \bibinfo {author} {\bibfnamefont {J.}~\bibnamefont
  {Vu{\v{c}}kovi{\'c}}},\ }\bibfield  {title} {\bibinfo {title} {Photonic
  quantum technologies},\ }\href@noop {} {\bibfield  {journal} {\bibinfo
  {journal} {Nature Photonics}\ }\textbf {\bibinfo {volume} {3}},\ \bibinfo
  {pages} {687} (\bibinfo {year} {2009})}\BibitemShut {NoStop}%
\bibitem [{\citenamefont {Flamini}\ \emph {et~al.}(2019)\citenamefont
  {Flamini}, \citenamefont {Spagnolo},\ and\ \citenamefont
  {Sciarrino}}]{Flamini:2019bp}%
  \BibitemOpen
  \bibfield  {author} {\bibinfo {author} {\bibfnamefont {F.}~\bibnamefont
  {Flamini}}, \bibinfo {author} {\bibfnamefont {N.}~\bibnamefont {Spagnolo}},\
  and\ \bibinfo {author} {\bibfnamefont {F.}~\bibnamefont {Sciarrino}},\
  }\bibfield  {title} {\bibinfo {title} {{Photonic quantum information
  processing: a review}},\ }\href@noop {} {\bibfield  {journal} {\bibinfo
  {journal} {Reports on Progress in Physics}\ }\textbf {\bibinfo {volume}
  {82}},\ \bibinfo {pages} {016001} (\bibinfo {year} {2019})}\BibitemShut
  {NoStop}%
\bibitem [{\citenamefont {Eisaman}\ \emph {et~al.}(2011)\citenamefont
  {Eisaman}, \citenamefont {Fan}, \citenamefont {Migdall},\ and\ \citenamefont
  {Polyakov}}]{eisaman2011invited}%
  \BibitemOpen
  \bibfield  {author} {\bibinfo {author} {\bibfnamefont {M.~D.}\ \bibnamefont
  {Eisaman}}, \bibinfo {author} {\bibfnamefont {J.}~\bibnamefont {Fan}},
  \bibinfo {author} {\bibfnamefont {A.}~\bibnamefont {Migdall}},\ and\ \bibinfo
  {author} {\bibfnamefont {S.~V.}\ \bibnamefont {Polyakov}},\ }\bibfield
  {title} {\bibinfo {title} {Invited review article: Single-photon sources and
  detectors},\ }\href@noop {} {\bibfield  {journal} {\bibinfo  {journal}
  {Review of scientific instruments}\ }\textbf {\bibinfo {volume} {82}},\
  \bibinfo {pages} {071101} (\bibinfo {year} {2011})}\BibitemShut {NoStop}%
\bibitem [{\citenamefont {Wang}\ \emph {et~al.}(2001)\citenamefont {Wang},
  \citenamefont {Hong},\ and\ \citenamefont {Friberg}}]{wang2001generation}%
  \BibitemOpen
  \bibfield  {author} {\bibinfo {author} {\bibfnamefont {L.}~\bibnamefont
  {Wang}}, \bibinfo {author} {\bibfnamefont {C.}~\bibnamefont {Hong}},\ and\
  \bibinfo {author} {\bibfnamefont {S.}~\bibnamefont {Friberg}},\ }\bibfield
  {title} {\bibinfo {title} {Generation of correlated photons via four-wave
  mixing in optical fibres},\ }\href@noop {} {\bibfield  {journal} {\bibinfo
  {journal} {Journal of optics B: Quantum and semiclassical optics}\ }\textbf
  {\bibinfo {volume} {3}},\ \bibinfo {pages} {346} (\bibinfo {year}
  {2001})}\BibitemShut {NoStop}%
\bibitem [{\citenamefont {Sharping}\ \emph {et~al.}(2001)\citenamefont
  {Sharping}, \citenamefont {Fiorentino},\ and\ \citenamefont
  {Kumar}}]{sharping2001observation}%
  \BibitemOpen
  \bibfield  {author} {\bibinfo {author} {\bibfnamefont {J.~E.}\ \bibnamefont
  {Sharping}}, \bibinfo {author} {\bibfnamefont {M.}~\bibnamefont
  {Fiorentino}},\ and\ \bibinfo {author} {\bibfnamefont {P.}~\bibnamefont
  {Kumar}},\ }\bibfield  {title} {\bibinfo {title} {Observation of
  twin-beam-type quantum correlation in optical fiber},\ }\href@noop {}
  {\bibfield  {journal} {\bibinfo  {journal} {Optics letters}\ }\textbf
  {\bibinfo {volume} {26}},\ \bibinfo {pages} {367} (\bibinfo {year}
  {2001})}\BibitemShut {NoStop}%
\bibitem [{\citenamefont {Fiorentino}\ \emph {et~al.}(2002)\citenamefont
  {Fiorentino}, \citenamefont {Voss}, \citenamefont {Sharping},\ and\
  \citenamefont {Kumar}}]{fiorentino2002all}%
  \BibitemOpen
  \bibfield  {author} {\bibinfo {author} {\bibfnamefont {M.}~\bibnamefont
  {Fiorentino}}, \bibinfo {author} {\bibfnamefont {P.~L.}\ \bibnamefont
  {Voss}}, \bibinfo {author} {\bibfnamefont {J.~E.}\ \bibnamefont {Sharping}},\
  and\ \bibinfo {author} {\bibfnamefont {P.}~\bibnamefont {Kumar}},\ }\bibfield
   {title} {\bibinfo {title} {All-fiber photon-pair source for quantum
  communications},\ }\href@noop {} {\bibfield  {journal} {\bibinfo  {journal}
  {IEEE Photonics Technology Letters}\ }\textbf {\bibinfo {volume} {14}},\
  \bibinfo {pages} {983} (\bibinfo {year} {2002})}\BibitemShut {NoStop}%
\bibitem [{\citenamefont {Takesue}\ and\ \citenamefont
  {Inoue}(2004)}]{takesue2004generation}%
  \BibitemOpen
  \bibfield  {author} {\bibinfo {author} {\bibfnamefont {H.}~\bibnamefont
  {Takesue}}\ and\ \bibinfo {author} {\bibfnamefont {K.}~\bibnamefont
  {Inoue}},\ }\bibfield  {title} {\bibinfo {title} {Generation of
  polarization-entangled photon pairs and violation of bell's inequality using
  spontaneous four-wave mixing in a fiber loop},\ }\href
  {https://doi.org/10.1103/PhysRevA.70.031802} {\bibfield  {journal} {\bibinfo
  {journal} {Phys. Rev. A}\ }\textbf {\bibinfo {volume} {70}},\ \bibinfo
  {pages} {031802} (\bibinfo {year} {2004})}\BibitemShut {NoStop}%
\bibitem [{\citenamefont {Li}\ \emph {et~al.}(2005)\citenamefont {Li},
  \citenamefont {Voss}, \citenamefont {Sharping},\ and\ \citenamefont
  {Kumar}}]{li2005optical}%
  \BibitemOpen
  \bibfield  {author} {\bibinfo {author} {\bibfnamefont {X.}~\bibnamefont
  {Li}}, \bibinfo {author} {\bibfnamefont {P.~L.}\ \bibnamefont {Voss}},
  \bibinfo {author} {\bibfnamefont {J.~E.}\ \bibnamefont {Sharping}},\ and\
  \bibinfo {author} {\bibfnamefont {P.}~\bibnamefont {Kumar}},\ }\bibfield
  {title} {\bibinfo {title} {Optical-fiber source of polarization-entangled
  photons in the 1550 nm telecom band},\ }\href@noop {} {\bibfield  {journal}
  {\bibinfo  {journal} {Physical review letters}\ }\textbf {\bibinfo {volume}
  {94}},\ \bibinfo {pages} {053601} (\bibinfo {year} {2005})}\BibitemShut
  {NoStop}%
\bibitem [{\citenamefont {Rarity}\ \emph {et~al.}(2005)\citenamefont {Rarity},
  \citenamefont {Fulconis}, \citenamefont {Duligall}, \citenamefont
  {Wadsworth},\ and\ \citenamefont {Russell}}]{rarity2005photonic}%
  \BibitemOpen
  \bibfield  {author} {\bibinfo {author} {\bibfnamefont {J.}~\bibnamefont
  {Rarity}}, \bibinfo {author} {\bibfnamefont {J.}~\bibnamefont {Fulconis}},
  \bibinfo {author} {\bibfnamefont {J.}~\bibnamefont {Duligall}}, \bibinfo
  {author} {\bibfnamefont {W.}~\bibnamefont {Wadsworth}},\ and\ \bibinfo
  {author} {\bibfnamefont {P.~S.~J.}\ \bibnamefont {Russell}},\ }\bibfield
  {title} {\bibinfo {title} {Photonic crystal fiber source of correlated photon
  pairs},\ }\href@noop {} {\bibfield  {journal} {\bibinfo  {journal} {Optics
  express}\ }\textbf {\bibinfo {volume} {13}},\ \bibinfo {pages} {534}
  (\bibinfo {year} {2005})}\BibitemShut {NoStop}%
\bibitem [{\citenamefont {Fan}\ \emph {et~al.}(2005)\citenamefont {Fan},
  \citenamefont {Migdall},\ and\ \citenamefont {Wang}}]{fan2005efficient}%
  \BibitemOpen
  \bibfield  {author} {\bibinfo {author} {\bibfnamefont {J.}~\bibnamefont
  {Fan}}, \bibinfo {author} {\bibfnamefont {A.}~\bibnamefont {Migdall}},\ and\
  \bibinfo {author} {\bibfnamefont {L.}~\bibnamefont {Wang}},\ }\bibfield
  {title} {\bibinfo {title} {Efficient generation of correlated photon pairs in
  a microstructure fiber},\ }\href@noop {} {\bibfield  {journal} {\bibinfo
  {journal} {Optics letters}\ }\textbf {\bibinfo {volume} {30}},\ \bibinfo
  {pages} {3368} (\bibinfo {year} {2005})}\BibitemShut {NoStop}%
\bibitem [{\citenamefont {Lin}\ \emph {et~al.}(2006)\citenamefont {Lin},
  \citenamefont {Yaman},\ and\ \citenamefont {Agrawal}}]{lin2006photon}%
  \BibitemOpen
  \bibfield  {author} {\bibinfo {author} {\bibfnamefont {Q.}~\bibnamefont
  {Lin}}, \bibinfo {author} {\bibfnamefont {F.}~\bibnamefont {Yaman}},\ and\
  \bibinfo {author} {\bibfnamefont {G.~P.}\ \bibnamefont {Agrawal}},\
  }\bibfield  {title} {\bibinfo {title} {Photon-pair generation by four-wave
  mixing in optical fibers},\ }\href@noop {} {\bibfield  {journal} {\bibinfo
  {journal} {Optics letters}\ }\textbf {\bibinfo {volume} {31}},\ \bibinfo
  {pages} {1286} (\bibinfo {year} {2006})}\BibitemShut {NoStop}%
\bibitem [{\citenamefont {Lee}\ \emph {et~al.}(2006)\citenamefont {Lee},
  \citenamefont {Chen}, \citenamefont {Liang}, \citenamefont {Li},
  \citenamefont {Voss},\ and\ \citenamefont {Kumar}}]{lee2006generation}%
  \BibitemOpen
  \bibfield  {author} {\bibinfo {author} {\bibfnamefont {K.~F.}\ \bibnamefont
  {Lee}}, \bibinfo {author} {\bibfnamefont {J.}~\bibnamefont {Chen}}, \bibinfo
  {author} {\bibfnamefont {C.}~\bibnamefont {Liang}}, \bibinfo {author}
  {\bibfnamefont {X.}~\bibnamefont {Li}}, \bibinfo {author} {\bibfnamefont
  {P.~L.}\ \bibnamefont {Voss}},\ and\ \bibinfo {author} {\bibfnamefont
  {P.}~\bibnamefont {Kumar}},\ }\bibfield  {title} {\bibinfo {title}
  {Generation of high-purity telecom-band entangled photon pairs in
  dispersion-shifted fiber},\ }\href@noop {} {\bibfield  {journal} {\bibinfo
  {journal} {Optics letters}\ }\textbf {\bibinfo {volume} {31}},\ \bibinfo
  {pages} {1905} (\bibinfo {year} {2006})}\BibitemShut {NoStop}%
\bibitem [{\citenamefont {Brambilla}(2010)}]{brambilla2010optical}%
  \BibitemOpen
  \bibfield  {author} {\bibinfo {author} {\bibfnamefont {G.}~\bibnamefont
  {Brambilla}},\ }\bibfield  {title} {\bibinfo {title} {Optical fibre nanowires
  and microwires: a review},\ }\href@noop {} {\bibfield  {journal} {\bibinfo
  {journal} {Journal of Optics}\ }\textbf {\bibinfo {volume} {12}},\ \bibinfo
  {pages} {043001} (\bibinfo {year} {2010})}\BibitemShut {NoStop}%
\bibitem [{\citenamefont {Tong}\ \emph {et~al.}(2012)\citenamefont {Tong},
  \citenamefont {Zi}, \citenamefont {Guo},\ and\ \citenamefont
  {Lou}}]{tong2012optical}%
  \BibitemOpen
  \bibfield  {author} {\bibinfo {author} {\bibfnamefont {L.}~\bibnamefont
  {Tong}}, \bibinfo {author} {\bibfnamefont {F.}~\bibnamefont {Zi}}, \bibinfo
  {author} {\bibfnamefont {X.}~\bibnamefont {Guo}},\ and\ \bibinfo {author}
  {\bibfnamefont {J.}~\bibnamefont {Lou}},\ }\bibfield  {title} {\bibinfo
  {title} {Optical microfibers and nanofibers: A tutorial},\ }\href@noop {}
  {\bibfield  {journal} {\bibinfo  {journal} {Optics Communications}\ }\textbf
  {\bibinfo {volume} {285}},\ \bibinfo {pages} {4641} (\bibinfo {year}
  {2012})}\BibitemShut {NoStop}%
\bibitem [{\citenamefont {Morrissey}\ \emph {et~al.}(2013)\citenamefont
  {Morrissey}, \citenamefont {Deasy}, \citenamefont {Frawley}, \citenamefont
  {Kumar}, \citenamefont {Prel}, \citenamefont {Russell}, \citenamefont
  {Truong},\ and\ \citenamefont {Nic~Chormaic}}]{morrissey2013spectroscopy}%
  \BibitemOpen
  \bibfield  {author} {\bibinfo {author} {\bibfnamefont {M.~J.}\ \bibnamefont
  {Morrissey}}, \bibinfo {author} {\bibfnamefont {K.}~\bibnamefont {Deasy}},
  \bibinfo {author} {\bibfnamefont {M.}~\bibnamefont {Frawley}}, \bibinfo
  {author} {\bibfnamefont {R.}~\bibnamefont {Kumar}}, \bibinfo {author}
  {\bibfnamefont {E.}~\bibnamefont {Prel}}, \bibinfo {author} {\bibfnamefont
  {L.}~\bibnamefont {Russell}}, \bibinfo {author} {\bibfnamefont {V.~G.}\
  \bibnamefont {Truong}},\ and\ \bibinfo {author} {\bibfnamefont
  {S.}~\bibnamefont {Nic~Chormaic}},\ }\bibfield  {title} {\bibinfo {title}
  {Spectroscopy, manipulation and trapping of neutral atoms, molecules, and
  other particles using optical nanofibers: a review},\ }\href@noop {}
  {\bibfield  {journal} {\bibinfo  {journal} {Sensors}\ }\textbf {\bibinfo
  {volume} {13}},\ \bibinfo {pages} {10449} (\bibinfo {year}
  {2013})}\BibitemShut {NoStop}%
\bibitem [{\citenamefont {Balykin}(2014)}]{balykin2014quantum}%
  \BibitemOpen
  \bibfield  {author} {\bibinfo {author} {\bibfnamefont {V.~I.}\ \bibnamefont
  {Balykin}},\ }\bibfield  {title} {\bibinfo {title} {Quantum manipulation of
  atoms and photons by using optical nanowaveguides},\ }\href@noop {}
  {\bibfield  {journal} {\bibinfo  {journal} {Uspekhi Fizicheskih Nauk}\
  }\textbf {\bibinfo {volume} {184}},\ \bibinfo {pages} {656} (\bibinfo {year}
  {2014})}\BibitemShut {NoStop}%
\bibitem [{\citenamefont {Nayak}\ \emph {et~al.}(2018)\citenamefont {Nayak},
  \citenamefont {Sadgrove}, \citenamefont {Yalla}, \citenamefont {Le~Kien},\
  and\ \citenamefont {Hakuta}}]{nayak2018nanofiber}%
  \BibitemOpen
  \bibfield  {author} {\bibinfo {author} {\bibfnamefont {K.~P.}\ \bibnamefont
  {Nayak}}, \bibinfo {author} {\bibfnamefont {M.}~\bibnamefont {Sadgrove}},
  \bibinfo {author} {\bibfnamefont {R.}~\bibnamefont {Yalla}}, \bibinfo
  {author} {\bibfnamefont {F.}~\bibnamefont {Le~Kien}},\ and\ \bibinfo {author}
  {\bibfnamefont {K.}~\bibnamefont {Hakuta}},\ }\bibfield  {title} {\bibinfo
  {title} {Nanofiber quantum photonics},\ }\href@noop {} {\bibfield  {journal}
  {\bibinfo  {journal} {Journal of Optics}\ }\textbf {\bibinfo {volume} {20}},\
  \bibinfo {pages} {073001} (\bibinfo {year} {2018})}\BibitemShut {NoStop}%
\bibitem [{\citenamefont {Chen}\ \emph {et~al.}(2013)\citenamefont {Chen},
  \citenamefont {Ding}, \citenamefont {Newson}, \citenamefont {Brambilla} \emph
  {et~al.}}]{chen2013review}%
  \BibitemOpen
  \bibfield  {author} {\bibinfo {author} {\bibfnamefont {G.~Y.}\ \bibnamefont
  {Chen}}, \bibinfo {author} {\bibfnamefont {M.}~\bibnamefont {Ding}}, \bibinfo
  {author} {\bibfnamefont {T.}~\bibnamefont {Newson}}, \bibinfo {author}
  {\bibfnamefont {G.}~\bibnamefont {Brambilla}}, \emph {et~al.},\ }\bibfield
  {title} {\bibinfo {title} {A review of microfiber and nanofiber based optical
  sensors},\ }\href@noop {} {\bibfield  {journal} {\bibinfo  {journal} {The
  Open Optics Journal}\ }\textbf {\bibinfo {volume} {7}} (\bibinfo {year}
  {2013})}\BibitemShut {NoStop}%
\bibitem [{\citenamefont {Spillane}\ \emph {et~al.}(2008)\citenamefont
  {Spillane}, \citenamefont {Pati}, \citenamefont {Salit}, \citenamefont
  {Hall}, \citenamefont {Kumar}, \citenamefont {Beausoleil},\ and\
  \citenamefont {Shahriar}}]{spillane2008observation}%
  \BibitemOpen
  \bibfield  {author} {\bibinfo {author} {\bibfnamefont {S.~M.}\ \bibnamefont
  {Spillane}}, \bibinfo {author} {\bibfnamefont {G.~S.}\ \bibnamefont {Pati}},
  \bibinfo {author} {\bibfnamefont {K.}~\bibnamefont {Salit}}, \bibinfo
  {author} {\bibfnamefont {M.}~\bibnamefont {Hall}}, \bibinfo {author}
  {\bibfnamefont {P.}~\bibnamefont {Kumar}}, \bibinfo {author} {\bibfnamefont
  {R.~G.}\ \bibnamefont {Beausoleil}},\ and\ \bibinfo {author} {\bibfnamefont
  {M.~S.}\ \bibnamefont {Shahriar}},\ }\bibfield  {title} {\bibinfo {title}
  {Observation of nonlinear optical interactions of ultralow levels of light in
  a tapered optical nanofiber embedded in a hot rubidium vapor},\ }\href
  {https://doi.org/10.1103/PhysRevLett.100.233602} {\bibfield  {journal}
  {\bibinfo  {journal} {Phys. Rev. Lett.}\ }\textbf {\bibinfo {volume} {100}},\
  \bibinfo {pages} {233602} (\bibinfo {year} {2008})}\BibitemShut {NoStop}%
\bibitem [{\citenamefont {Le~Kien}\ \emph {et~al.}(2004)\citenamefont
  {Le~Kien}, \citenamefont {Balykin},\ and\ \citenamefont
  {Hakuta}}]{le2004atom}%
  \BibitemOpen
  \bibfield  {author} {\bibinfo {author} {\bibfnamefont {F.}~\bibnamefont
  {Le~Kien}}, \bibinfo {author} {\bibfnamefont {V.~I.}\ \bibnamefont
  {Balykin}},\ and\ \bibinfo {author} {\bibfnamefont {K.}~\bibnamefont
  {Hakuta}},\ }\bibfield  {title} {\bibinfo {title} {Atom trap and waveguide
  using a two-color evanescent light field around a subwavelength-diameter
  optical fiber},\ }\href {https://doi.org/10.1103/PhysRevA.70.063403}
  {\bibfield  {journal} {\bibinfo  {journal} {Phys. Rev. A}\ }\textbf {\bibinfo
  {volume} {70}},\ \bibinfo {pages} {063403} (\bibinfo {year}
  {2004})}\BibitemShut {NoStop}%
\bibitem [{\citenamefont {Le~Kien}\ \emph {et~al.}(2005)\citenamefont
  {Le~Kien}, \citenamefont {Dutta~Gupta}, \citenamefont {Balykin},\ and\
  \citenamefont {Hakuta}}]{le2005spontaneous}%
  \BibitemOpen
  \bibfield  {author} {\bibinfo {author} {\bibfnamefont {F.}~\bibnamefont
  {Le~Kien}}, \bibinfo {author} {\bibfnamefont {S.}~\bibnamefont
  {Dutta~Gupta}}, \bibinfo {author} {\bibfnamefont {V.~I.}\ \bibnamefont
  {Balykin}},\ and\ \bibinfo {author} {\bibfnamefont {K.}~\bibnamefont
  {Hakuta}},\ }\bibfield  {title} {\bibinfo {title} {Publisher's note:
  Spontaneous emission of a cesium atom near a nanofiber: Efficient coupling of
  light to guided modes [phys. rev. a 72, 032509 (2005)]},\ }\href
  {https://doi.org/10.1103/PhysRevA.72.049904} {\bibfield  {journal} {\bibinfo
  {journal} {Phys. Rev. A}\ }\textbf {\bibinfo {volume} {72}},\ \bibinfo
  {pages} {049904} (\bibinfo {year} {2005})}\BibitemShut {NoStop}%
\bibitem [{\citenamefont {Cui}\ \emph {et~al.}(2013)\citenamefont {Cui},
  \citenamefont {Li}, \citenamefont {Guo}, \citenamefont {Li}, \citenamefont
  {Xu}, \citenamefont {Wang},\ and\ \citenamefont {Fang}}]{cui2013generation}%
  \BibitemOpen
  \bibfield  {author} {\bibinfo {author} {\bibfnamefont {L.}~\bibnamefont
  {Cui}}, \bibinfo {author} {\bibfnamefont {X.}~\bibnamefont {Li}}, \bibinfo
  {author} {\bibfnamefont {C.}~\bibnamefont {Guo}}, \bibinfo {author}
  {\bibfnamefont {Y.}~\bibnamefont {Li}}, \bibinfo {author} {\bibfnamefont
  {Z.}~\bibnamefont {Xu}}, \bibinfo {author} {\bibfnamefont {L.}~\bibnamefont
  {Wang}},\ and\ \bibinfo {author} {\bibfnamefont {W.}~\bibnamefont {Fang}},\
  }\bibfield  {title} {\bibinfo {title} {Generation of correlated photon pairs
  in micro/nano-fibers},\ }\href@noop {} {\bibfield  {journal} {\bibinfo
  {journal} {Optics letters}\ }\textbf {\bibinfo {volume} {38}},\ \bibinfo
  {pages} {5063} (\bibinfo {year} {2013})}\BibitemShut {NoStop}%
\bibitem [{\citenamefont {Su}\ \emph {et~al.}(2018)\citenamefont {Su},
  \citenamefont {Cui}, \citenamefont {Li},\ and\ \citenamefont
  {Li}}]{su2018micro}%
  \BibitemOpen
  \bibfield  {author} {\bibinfo {author} {\bibfnamefont {J.}~\bibnamefont
  {Su}}, \bibinfo {author} {\bibfnamefont {L.}~\bibnamefont {Cui}}, \bibinfo
  {author} {\bibfnamefont {Y.}~\bibnamefont {Li}},\ and\ \bibinfo {author}
  {\bibfnamefont {X.}~\bibnamefont {Li}},\ }\bibfield  {title} {\bibinfo
  {title} {Micro/nano-fiber-based source of heralded single photons at the
  telecom band},\ }\href@noop {} {\bibfield  {journal} {\bibinfo  {journal}
  {Chinese Optics Letters}\ }\textbf {\bibinfo {volume} {16}},\ \bibinfo
  {pages} {041903} (\bibinfo {year} {2018})}\BibitemShut {NoStop}%
\bibitem [{\citenamefont {Kim}\ \emph {et~al.}(2019)\citenamefont {Kim},
  \citenamefont {Ihn}, \citenamefont {Kim},\ and\ \citenamefont
  {Shin}}]{kim2019photon}%
  \BibitemOpen
  \bibfield  {author} {\bibinfo {author} {\bibfnamefont {J.-H.}\ \bibnamefont
  {Kim}}, \bibinfo {author} {\bibfnamefont {Y.~S.}\ \bibnamefont {Ihn}},
  \bibinfo {author} {\bibfnamefont {Y.-H.}\ \bibnamefont {Kim}},\ and\ \bibinfo
  {author} {\bibfnamefont {H.}~\bibnamefont {Shin}},\ }\bibfield  {title}
  {\bibinfo {title} {Photon-pair source working in a silicon-based detector
  wavelength range using tapered micro/nanofibers},\ }\href@noop {} {\bibfield
  {journal} {\bibinfo  {journal} {Optics letters}\ }\textbf {\bibinfo {volume}
  {44}},\ \bibinfo {pages} {447} (\bibinfo {year} {2019})}\BibitemShut
  {NoStop}%
\bibitem [{\citenamefont {Meyer-Scott}\ \emph {et~al.}(2015)\citenamefont
  {Meyer-Scott}, \citenamefont {Dot}, \citenamefont {Ahmad}, \citenamefont
  {Li}, \citenamefont {Rochette},\ and\ \citenamefont
  {Jennewein}}]{meyer2015power}%
  \BibitemOpen
  \bibfield  {author} {\bibinfo {author} {\bibfnamefont {E.}~\bibnamefont
  {Meyer-Scott}}, \bibinfo {author} {\bibfnamefont {A.}~\bibnamefont {Dot}},
  \bibinfo {author} {\bibfnamefont {R.}~\bibnamefont {Ahmad}}, \bibinfo
  {author} {\bibfnamefont {L.}~\bibnamefont {Li}}, \bibinfo {author}
  {\bibfnamefont {M.}~\bibnamefont {Rochette}},\ and\ \bibinfo {author}
  {\bibfnamefont {T.}~\bibnamefont {Jennewein}},\ }\bibfield  {title} {\bibinfo
  {title} {Power-efficient production of photon pairs in a tapered chalcogenide
  microwire},\ }\href@noop {} {\bibfield  {journal} {\bibinfo  {journal}
  {Applied Physics Letters}\ }\textbf {\bibinfo {volume} {106}},\ \bibinfo
  {pages} {081111} (\bibinfo {year} {2015})}\BibitemShut {NoStop}%
\bibitem [{\citenamefont {Ward}\ \emph {et~al.}(2014)\citenamefont {Ward},
  \citenamefont {Maimaiti}, \citenamefont {Le},\ and\ \citenamefont
  {Chormaic}}]{ward2014contributed}%
  \BibitemOpen
  \bibfield  {author} {\bibinfo {author} {\bibfnamefont {J.}~\bibnamefont
  {Ward}}, \bibinfo {author} {\bibfnamefont {A.}~\bibnamefont {Maimaiti}},
  \bibinfo {author} {\bibfnamefont {V.~H.}\ \bibnamefont {Le}},\ and\ \bibinfo
  {author} {\bibfnamefont {S.~N.}\ \bibnamefont {Chormaic}},\ }\bibfield
  {title} {\bibinfo {title} {Contributed review: Optical micro-and nanofiber
  pulling rig},\ }\href@noop {} {\bibfield  {journal} {\bibinfo  {journal}
  {Review of Scientific Instruments}\ }\textbf {\bibinfo {volume} {85}},\
  \bibinfo {pages} {111501} (\bibinfo {year} {2014})}\BibitemShut {NoStop}%
\bibitem [{\citenamefont {Garay-Palmett}\ \emph {et~al.}(2010)\citenamefont
  {Garay-Palmett}, \citenamefont {U’Ren},\ and\ \citenamefont
  {Rangel-Rojo}}]{garay2010conversion}%
  \BibitemOpen
  \bibfield  {author} {\bibinfo {author} {\bibfnamefont {K.}~\bibnamefont
  {Garay-Palmett}}, \bibinfo {author} {\bibfnamefont {A.~B.}\ \bibnamefont
  {U’Ren}},\ and\ \bibinfo {author} {\bibfnamefont {R.}~\bibnamefont
  {Rangel-Rojo}},\ }\bibfield  {title} {\bibinfo {title} {Conversion efficiency
  in the process of copolarized spontaneous four-wave mixing},\ }\href@noop {}
  {\bibfield  {journal} {\bibinfo  {journal} {Physical Review A}\ }\textbf
  {\bibinfo {volume} {82}},\ \bibinfo {pages} {043809} (\bibinfo {year}
  {2010})}\BibitemShut {NoStop}%
\bibitem [{\citenamefont {Katsenelenbaum}(1961)}]{Katsenelenbaum}%
  \BibitemOpen
  \bibfield  {author} {\bibinfo {author} {\bibfnamefont {B.}~\bibnamefont
  {Katsenelenbaum}},\ }\href@noop {} {\emph {\bibinfo {title} {The theory of
  irregular waveguides with slowly varying parameters [In Russian]}}}\
  (\bibinfo  {publisher} {USSR Academy of Sciences Publisher, Moscow},\
  \bibinfo {year} {1961})\ \bibinfo {note} {in Russian}\BibitemShut {NoStop}%
\bibitem [{\citenamefont {Mosley}\ \emph {et~al.}(2008)\citenamefont {Mosley},
  \citenamefont {Lundeen}, \citenamefont {Smith}, \citenamefont {Wasylczyk},
  \citenamefont {U’Ren}, \citenamefont {Silberhorn},\ and\ \citenamefont
  {Walmsley}}]{mosley2008heralded}%
  \BibitemOpen
  \bibfield  {author} {\bibinfo {author} {\bibfnamefont {P.~J.}\ \bibnamefont
  {Mosley}}, \bibinfo {author} {\bibfnamefont {J.~S.}\ \bibnamefont {Lundeen}},
  \bibinfo {author} {\bibfnamefont {B.~J.}\ \bibnamefont {Smith}}, \bibinfo
  {author} {\bibfnamefont {P.}~\bibnamefont {Wasylczyk}}, \bibinfo {author}
  {\bibfnamefont {A.~B.}\ \bibnamefont {U’Ren}}, \bibinfo {author}
  {\bibfnamefont {C.}~\bibnamefont {Silberhorn}},\ and\ \bibinfo {author}
  {\bibfnamefont {I.~A.}\ \bibnamefont {Walmsley}},\ }\bibfield  {title}
  {\bibinfo {title} {Heralded generation of ultrafast single photons in pure
  quantum states},\ }\href@noop {} {\bibfield  {journal} {\bibinfo  {journal}
  {Physical Review Letters}\ }\textbf {\bibinfo {volume} {100}},\ \bibinfo
  {pages} {133601} (\bibinfo {year} {2008})}\BibitemShut {NoStop}%
\bibitem [{\citenamefont {Spring}\ \emph {et~al.}(2013)\citenamefont {Spring},
  \citenamefont {Salter}, \citenamefont {Metcalf}, \citenamefont {Humphreys},
  \citenamefont {Moore}, \citenamefont {Thomas-Peter}, \citenamefont
  {Barbieri}, \citenamefont {Jin}, \citenamefont {Langford}, \citenamefont
  {Kolthammer} \emph {et~al.}}]{spring2013chip}%
  \BibitemOpen
  \bibfield  {author} {\bibinfo {author} {\bibfnamefont {J.~B.}\ \bibnamefont
  {Spring}}, \bibinfo {author} {\bibfnamefont {P.~S.}\ \bibnamefont {Salter}},
  \bibinfo {author} {\bibfnamefont {B.~J.}\ \bibnamefont {Metcalf}}, \bibinfo
  {author} {\bibfnamefont {P.~C.}\ \bibnamefont {Humphreys}}, \bibinfo {author}
  {\bibfnamefont {M.}~\bibnamefont {Moore}}, \bibinfo {author} {\bibfnamefont
  {N.}~\bibnamefont {Thomas-Peter}}, \bibinfo {author} {\bibfnamefont
  {M.}~\bibnamefont {Barbieri}}, \bibinfo {author} {\bibfnamefont {X.-M.}\
  \bibnamefont {Jin}}, \bibinfo {author} {\bibfnamefont {N.~K.}\ \bibnamefont
  {Langford}}, \bibinfo {author} {\bibfnamefont {W.~S.}\ \bibnamefont
  {Kolthammer}}, \emph {et~al.},\ }\bibfield  {title} {\bibinfo {title}
  {On-chip low loss heralded source of pure single photons},\ }\href@noop {}
  {\bibfield  {journal} {\bibinfo  {journal} {Optics express}\ }\textbf
  {\bibinfo {volume} {21}},\ \bibinfo {pages} {13522} (\bibinfo {year}
  {2013})}\BibitemShut {NoStop}%
\bibitem [{\citenamefont {Agrawal}(2012)}]{agrawal2012nonlinear}%
  \BibitemOpen
  \bibfield  {author} {\bibinfo {author} {\bibfnamefont {G.}~\bibnamefont
  {Agrawal}},\ }\href@noop {} {\emph {\bibinfo {title} {Nonlinear fiber
  optics}}}\ (\bibinfo  {publisher} {Academic Press, New York},\ \bibinfo
  {year} {2012})\BibitemShut {NoStop}%
\bibitem [{\citenamefont {Garay-Palmett}\ \emph {et~al.}(2011)\citenamefont
  {Garay-Palmett}, \citenamefont {Corona},\ and\ \citenamefont
  {U’Ren}}]{garay2011spontaneous}%
  \BibitemOpen
  \bibfield  {author} {\bibinfo {author} {\bibfnamefont {K.}~\bibnamefont
  {Garay-Palmett}}, \bibinfo {author} {\bibfnamefont {M.}~\bibnamefont
  {Corona}},\ and\ \bibinfo {author} {\bibfnamefont {A.}~\bibnamefont
  {U’Ren}},\ }\bibfield  {title} {\bibinfo {title} {Spontaneous parametric
  processes in optical fibers: a comparison},\ }\href@noop {} {\bibfield
  {journal} {\bibinfo  {journal} {Rev. Mex. F{\i}s. S}\ }\textbf {\bibinfo
  {volume} {57}},\ \bibinfo {pages} {6} (\bibinfo {year} {2011})}\BibitemShut
  {NoStop}%
\bibitem [{\citenamefont {Fasel}\ \emph {et~al.}(2004)\citenamefont {Fasel},
  \citenamefont {Alibart}, \citenamefont {Tanzilli}, \citenamefont {Baldi},
  \citenamefont {Beveratos}, \citenamefont {Gisin},\ and\ \citenamefont
  {Zbinden}}]{fasel2004high}%
  \BibitemOpen
  \bibfield  {author} {\bibinfo {author} {\bibfnamefont {S.}~\bibnamefont
  {Fasel}}, \bibinfo {author} {\bibfnamefont {O.}~\bibnamefont {Alibart}},
  \bibinfo {author} {\bibfnamefont {S.}~\bibnamefont {Tanzilli}}, \bibinfo
  {author} {\bibfnamefont {P.}~\bibnamefont {Baldi}}, \bibinfo {author}
  {\bibfnamefont {A.}~\bibnamefont {Beveratos}}, \bibinfo {author}
  {\bibfnamefont {N.}~\bibnamefont {Gisin}},\ and\ \bibinfo {author}
  {\bibfnamefont {H.}~\bibnamefont {Zbinden}},\ }\bibfield  {title} {\bibinfo
  {title} {High-quality asynchronous heralded single-photon source at telecom
  wavelength},\ }\href@noop {} {\bibfield  {journal} {\bibinfo  {journal} {New
  Journal of Physics}\ }\textbf {\bibinfo {volume} {6}},\ \bibinfo {pages}
  {163} (\bibinfo {year} {2004})}\BibitemShut {NoStop}%
\bibitem [{\citenamefont {Riel{\"a}nder}\ \emph {et~al.}(2016)\citenamefont
  {Riel{\"a}nder}, \citenamefont {Lenhard}, \citenamefont {Mazzera},\ and\
  \citenamefont {de~Riedmatten}}]{rielander2016cavity}%
  \BibitemOpen
  \bibfield  {author} {\bibinfo {author} {\bibfnamefont {D.}~\bibnamefont
  {Riel{\"a}nder}}, \bibinfo {author} {\bibfnamefont {A.}~\bibnamefont
  {Lenhard}}, \bibinfo {author} {\bibfnamefont {M.}~\bibnamefont {Mazzera}},\
  and\ \bibinfo {author} {\bibfnamefont {H.}~\bibnamefont {de~Riedmatten}},\
  }\bibfield  {title} {\bibinfo {title} {Cavity enhanced telecom heralded
  single photons for spin-wave solid state quantum memories},\ }\href@noop {}
  {\bibfield  {journal} {\bibinfo  {journal} {New Journal of Physics}\ }\textbf
  {\bibinfo {volume} {18}} (\bibinfo {year} {2016})}\BibitemShut {NoStop}%
\bibitem [{\citenamefont {Seri}\ \emph {et~al.}(2019)\citenamefont {Seri},
  \citenamefont {Lago-Rivera}, \citenamefont {Lenhard}, \citenamefont
  {Corrielli}, \citenamefont {Osellame}, \citenamefont {Mazzera},\ and\
  \citenamefont {de~Riedmatten}}]{seri2019quantum}%
  \BibitemOpen
  \bibfield  {author} {\bibinfo {author} {\bibfnamefont {A.}~\bibnamefont
  {Seri}}, \bibinfo {author} {\bibfnamefont {D.}~\bibnamefont {Lago-Rivera}},
  \bibinfo {author} {\bibfnamefont {A.}~\bibnamefont {Lenhard}}, \bibinfo
  {author} {\bibfnamefont {G.}~\bibnamefont {Corrielli}}, \bibinfo {author}
  {\bibfnamefont {R.}~\bibnamefont {Osellame}}, \bibinfo {author}
  {\bibfnamefont {M.}~\bibnamefont {Mazzera}},\ and\ \bibinfo {author}
  {\bibfnamefont {H.}~\bibnamefont {de~Riedmatten}},\ }\bibfield  {title}
  {\bibinfo {title} {Quantum storage of frequency-multiplexed heralded single
  photons},\ }\href@noop {} {\bibfield  {journal} {\bibinfo  {journal} {arXiv
  preprint arXiv:1902.06657}\ } (\bibinfo {year} {2019})}\BibitemShut {NoStop}%
\bibitem [{\citenamefont {Seri}\ \emph {et~al.}(2018)\citenamefont {Seri},
  \citenamefont {Corrielli}, \citenamefont {Lago-Rivera}, \citenamefont
  {Lenhard}, \citenamefont {de~Riedmatten}, \citenamefont {Osellame},\ and\
  \citenamefont {Mazzera}}]{seri2018laser}%
  \BibitemOpen
  \bibfield  {author} {\bibinfo {author} {\bibfnamefont {A.}~\bibnamefont
  {Seri}}, \bibinfo {author} {\bibfnamefont {G.}~\bibnamefont {Corrielli}},
  \bibinfo {author} {\bibfnamefont {D.}~\bibnamefont {Lago-Rivera}}, \bibinfo
  {author} {\bibfnamefont {A.}~\bibnamefont {Lenhard}}, \bibinfo {author}
  {\bibfnamefont {H.}~\bibnamefont {de~Riedmatten}}, \bibinfo {author}
  {\bibfnamefont {R.}~\bibnamefont {Osellame}},\ and\ \bibinfo {author}
  {\bibfnamefont {M.}~\bibnamefont {Mazzera}},\ }\bibfield  {title} {\bibinfo
  {title} {Laser-written integrated platform for quantum storage of heralded
  single photons},\ }\href@noop {} {\bibfield  {journal} {\bibinfo  {journal}
  {Optica}\ }\textbf {\bibinfo {volume} {5}},\ \bibinfo {pages} {934} (\bibinfo
  {year} {2018})}\BibitemShut {NoStop}%
\bibitem [{\citenamefont {Lenhard}\ \emph {et~al.}(2017)\citenamefont
  {Lenhard}, \citenamefont {Brito}, \citenamefont {Bock}, \citenamefont
  {Becher},\ and\ \citenamefont {Eschner}}]{lenhard2017coherence}%
  \BibitemOpen
  \bibfield  {author} {\bibinfo {author} {\bibfnamefont {A.}~\bibnamefont
  {Lenhard}}, \bibinfo {author} {\bibfnamefont {J.}~\bibnamefont {Brito}},
  \bibinfo {author} {\bibfnamefont {M.}~\bibnamefont {Bock}}, \bibinfo {author}
  {\bibfnamefont {C.}~\bibnamefont {Becher}},\ and\ \bibinfo {author}
  {\bibfnamefont {J.}~\bibnamefont {Eschner}},\ }\bibfield  {title} {\bibinfo
  {title} {Coherence and entanglement preservation of frequency-converted
  heralded single photons},\ }\href@noop {} {\bibfield  {journal} {\bibinfo
  {journal} {Optics express}\ }\textbf {\bibinfo {volume} {25}},\ \bibinfo
  {pages} {11187} (\bibinfo {year} {2017})}\BibitemShut {NoStop}%
\bibitem [{\citenamefont {Bock}\ \emph {et~al.}(2016)\citenamefont {Bock},
  \citenamefont {Lenhard}, \citenamefont {Chunnilall},\ and\ \citenamefont
  {Becher}}]{bock2016highly}%
  \BibitemOpen
  \bibfield  {author} {\bibinfo {author} {\bibfnamefont {M.}~\bibnamefont
  {Bock}}, \bibinfo {author} {\bibfnamefont {A.}~\bibnamefont {Lenhard}},
  \bibinfo {author} {\bibfnamefont {C.}~\bibnamefont {Chunnilall}},\ and\
  \bibinfo {author} {\bibfnamefont {C.}~\bibnamefont {Becher}},\ }\bibfield
  {title} {\bibinfo {title} {Highly efficient heralded single-photon source for
  telecom wavelengths based on a ppln waveguide},\ }\href@noop {} {\bibfield
  {journal} {\bibinfo  {journal} {Optics express}\ }\textbf {\bibinfo {volume}
  {24}},\ \bibinfo {pages} {23992} (\bibinfo {year} {2016})}\BibitemShut
  {NoStop}%
\end{thebibliography}%

\end{document}